\RequirePackage{fix-cm}
\documentclass[twocolumn,epjc3]{svjour3}  

\smartqed  						
\RequirePackage{graphicx}
\RequirePackage{mathptmx}      	
\RequirePackage[numbers,sort&compress]{natbib}
\RequirePackage[colorlinks,citecolor=blue,urlcolor=blue,linkcolor=blue]{hyperref}

\usepackage{arydshln}
\newcommand{\sigmav}{\langle \sigma_{\mathrm{A}} v\rangle}
\newcommand{\GeV}{\mathrm{GeV}}
\newcommand{\cm}{\mathrm{cm}}
\newcommand{\TeV}{\mathrm{TeV}}
\newcommand{\EeV}{\mathrm{EeV}}
\newcommand{\bit}{\begin{itemize}}
\newcommand{\eit}{\end{itemize}}

\journalname{Eur. Phys. J. C}

\begin{document}
\title{Multipole analysis of IceCube data to search for dark matter accumulated in the Galactic halo}

\onecolumn   
\author{IceCube Collaboration: M.~G.~Aartsen\thanksref{Adelaide}
\and M.~Ackermann\thanksref{Zeuthen}
\and J.~Adams\thanksref{Christchurch}
\and J.~A.~Aguilar\thanksref{Geneva}
\and M.~Ahlers\thanksref{MadisonPAC}
\and M.~Ahrens\thanksref{StockholmOKC}
\and D.~Altmann\thanksref{Erlangen}
\and T.~Anderson\thanksref{PennPhys}
\and C.~Arguelles\thanksref{MadisonPAC}
\and T.~C.~Arlen\thanksref{PennPhys}
% ...
}
\authorrunning{IceCube Collaboration}
\thankstext{a}{Corresponding author: reimann@physik.rwth-aachen.de}
\thankstext{b}{Earthquake Research Institute, University of Tokyo, Bunkyo, Tokyo 113-0032, Japan}
\thankstext{c}{NASA Goddard Space Flight Center, Greenbelt, MD 20771, USA}
\institute{
% ...
School of Chemistry \& Physics, University of Adelaide, Adelaide SA, 5005 Australia \label{Adelaide}
% ...
\and Dept.~of Physics and Astronomy, University of Canterbury, Private Bag 4800, Christchurch, New Zealand \label{Christchurch}
% ...
\and Erlangen Centre for Astroparticle Physics, Friedrich-Alexander-Universit\"at Erlangen-N\"urnberg, D-91058 Erlangen, Germany \label{Erlangen}
\and D\'epartement de physique nucl\'eaire et corpusculaire, Universit\'e de Gen\`eve, CH-1211 Gen\`eve, Switzerland \label{Geneva}
% ...
\and Dept.~of Physics and Wisconsin IceCube Particle Astrophysics Center, University of Wisconsin, Madison, WI 53706, USA \label{MadisonPAC}
% ...
\and Oskar Klein Centre and Dept.~of Physics, Stockholm University, SE-10691 Stockholm, Sweden \label{StockholmOKC}
% ...
\and Dept.~of Physics, Pennsylvania State University, University Park, PA 16802, USA \label{PennPhys}
% ...
\and DESY, D-15735 Zeuthen, Germany \label{Zeuthen}
} 


\date{Received: date / Accepted: date} 		
\maketitle
\twocolumn
\hyphenation{IceCube}
\sloppy


\begin{abstract}
	Dark matter which is bound in the Galactic halo might 
	self-annihilate and produce a flux of stable final state particles, 
	e.g. high energy neutrinos. These neutrinos can be detected with 
	IceCube, a cubic-kilometer sized Cherenkov detector. Given 
	IceCube's large field of view, a characteristic anisotropy of 
	the additional neutrino flux is expected. In this paper we 
	describe a multipole method to search for such a large-scale 
	anisotropy in IceCube data. This method uses the 
	expansion coefficients of a multipole expansion of neutrino 
	arrival directions and incorporates signal-specific weights for 
	each expansion coefficient. We apply the technique to a 
	high-purity muon neutrino sample from the Northern Hemisphere. 
	The final result is compatible with the null-hypothesis. As no 
	signal was observed, we present limits on the self-annihilation 
	cross-section averaged over the relative velocity distribution $\sigmav$ down to $1.9\cdot 
	10^{-23}\,\textrm{cm}^3\textrm{s}^{-1}$ for a dark matter particle 
	mass of $700\,\GeV$ to $1000\,\GeV$ and direct annihilation into $\nu\bar{\nu}$. 
	The resulting exclusion 
	limits come close to exclusion limits from 
	$\gamma$-ray experiments, that focus on the outer Galactic halo, for high dark matter masses of a few 
	TeV and hard annihilation channels.

\end{abstract}


\section{Introduction}
\label{sec:intro}

There is compelling evidence for dark matter, e.g. from cosmic 
microwave background anisotropies, 
large-scale structure formation, galaxy rotation-curves, and other astrophysical observations~\cite{wmap,Planck,Bertone:2004pz}. 
Despite this evidence, DM can not be (fully) explained by standard model
particles, and its nature remains unknown~\cite{Bertone:2004pz}.
Many theories, e.g. supersymmetry or extra dimensions, provide suitable 
candidates for dark matter~\cite{Bertone:2004pz}. The generic candidate for 
dark matter is a weakly interacting massive particle (WIMP) with a mass of a
few $\GeV$ up to several hundred $\TeV$~\cite{WIMPMass,WIMPMass2}. 
Assuming that WIMPs interact at the scale of the weak 
force and were produced in the early universe in thermal equilibrium, 
the freeze-out of WIMPs leads to an expected dark matter abundance that is 
compatible with current estimates~\cite{Planck}. 

The density of WIMPs gravitationally trapped as dark halos in galaxies can be high
enough that their pair-wise annihilation rate is not negligible. The
final-state products of the annihilations decay to stable standard model particles, i.e., photons,
protons, electrons or neutrinos, and, therefore, an observable flux of these particles
could provide indirect evidence for dark matter.
While charged cosmic rays are deflected by magnetic 
fields and photons have a large astrophysical foreground, astrophysical neutrinos 
from dark matter annihilation do not interact with inter-stellar matter 
and would point back to their origin. 
In certain models, neutrinos can also be produced directly~\cite{lindner_enhancing_2010}, giving a monochromatic neutrino signal
that would be a golden channel for neutrino telescopes.
 
Observations of 
an excess in the positron to electron ratio by PAMELA~\cite{Adriani:2008zr}, 
that was confirmed by FERMI~\cite{PhysRevLett.108.011103} and 
AMS-02~\cite{PhysRevLett.110.141102, PhysRevLett.113.121101}, may hint to dark matter in the GeV-TeV region. 
The nature of the positron signal is extremely difficult to interpret due 
to the complex propagation of electrons and positrons in the Galactic magnetic fields. 
The observation can also be explained by nearby astrophysical sources like pulsars~\cite{Moskalenko:1997gh} or supernova remnants~\cite{Yuksel:2008rf}. 
However, if the positron excess is interpreted as originating from dark matter, 
leptophilic dark matter~\cite{Meade:2009iu, Cirelli:2008pk} is favored, 
with cross-sections in the range $10^{-24}\,\mathrm{cm}^3\mathrm{s}^{-1}$ to $10^{-21}\,\mathrm{cm}^3\mathrm{s}^{-1}$, 
partly within the sensitivity reach of the analysis presented here.

As mentioned above, the annihilation rate is significantly enhanced in regions where
DM might have been gravitationally accumulated, since the annihilation rate scales with the
square of the density. In particular, massive bodies like the Sun~\cite{PhysRevD.85.042002}, the Earth~\cite{ahrens_limits_2002}, the Galactic Center~\cite{the_icecube_collaboration_search_2012, newGCAnalysis} or dwarf galaxies and galaxy clusters ~\cite{aartsen_icecube_2013}, are good candidates
to search for a neutrino flux from DM annihilations. Furthermore, and due to the expected shape
of the dark halo around the Milky Way, annihilations in the halo would produce a diffuse flux
of neutrinos with a characteristic large-scale structure~\cite{IC22}, depending on the assumed DM
density distribution. While searches for a neutrino flux from the annihilation of
DM captured in massive bodies are sensitive to the spin-dependent and spin-independent
DM-nucleon cross -section, the Galactic and extra-galactic flux depends on the self-annihilation
cross section~\cite{Bertone:2004pz}.

In this paper we present a multipole method to search for a characteristic 
anisotropic flux of neutrinos from dark matter annihilation in the Galactic 
halo. The method is based on a multipole expansion of the sky map of 
arrival directions and an optimized test statistic using the 
expansion coefficients. 
This method provides the opportunity to reduce
the influence of systematic uncertainties in the result, which arise from systematic
uncertainties on the zenith dependent 
acceptance and zenith dependent atmospheric neutrino flux. A 
large-scale anisotropy as seen by~\cite{MilagroCRAnisotropy,sveshnikova_spectrum_2013,abbasi_measurement_2010,bartoli_medium_2013} in cosmic-rays, is 
expected in the atmospheric neutrino flux. However this anisotropy 
is very small so that it is just an effect of few percent compared to our 
sensitivity on neutrinos from dark matter annihilations. 

This paper is organized as follows: In Section~\ref{sec:detector} 
the IceCube Neutrino Observatory is introduced. 
Section~\ref{sec:theo} gives the theoretical expected flux from dark matter 
annihilation with neutrinos as final state.  Section~\ref{sec:sample} 
gives a short overview of the data sample used and the simulation of 
pseudo-experiments. In Section~\ref{sec:method} the multipole 
analysis technique is introduced. The sensitivity of this 
analysis is given in Section~\ref{sec:sensitivity}. 
Section~\ref{sec:systematics} addresses systematic uncertainties. In 
Section~\ref{sec:results} and Section~\ref{sec:discussion} the 
experimental result is presented and discussed, while in
Section~\ref{sec:conclusion} we present our conclusions.


\section{The IceCube Neutrino Observatory}
\label{sec:detector}

\begin{figure}
    \includegraphics[width=1.\linewidth]{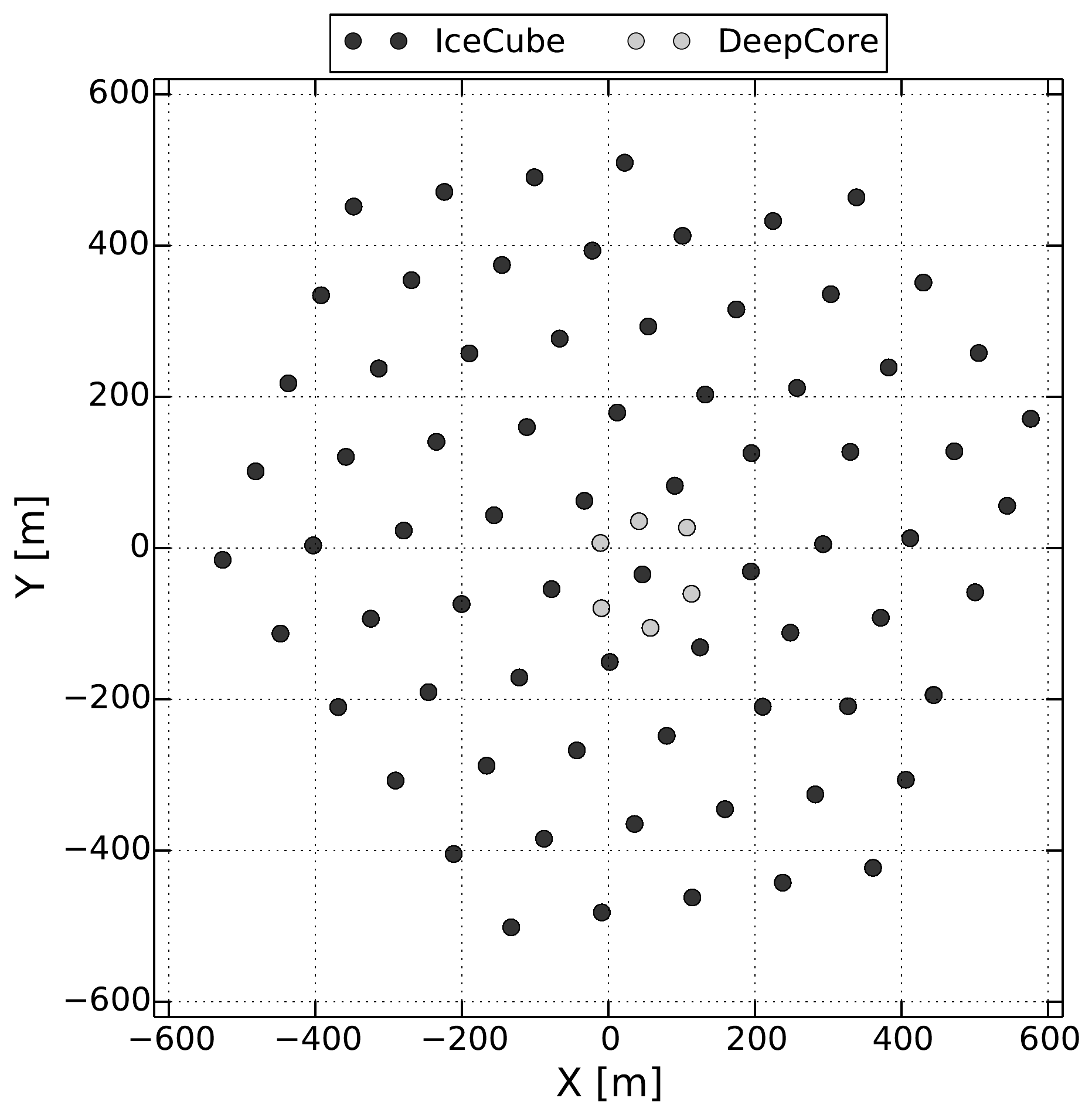} 
	\caption{Footprint of the IceCube detector in its 79-string 
	configuration, that was taking data from June 2010 to May 2011. 
	Shown is the position of the strings, where standard IceCube 
	strings are marked in dark gray, and DeepCore strings with a smaller DOM 
	spacing are marked in light gray.}
    \label{fig:detector}
\end{figure}

IceCube is a cubic-kilometer Cherenkov neutrino detector located at the geographic 
South Pole~\cite{IceCubeDetector}. 
When a neutrino interacts with the clear Antarctic ice, secondary leptons and hadrons are produced.
These relativistic secondary particles produce Cherenkov light which is 
detected by Digital Optical Modules (DOMs) that contain a 
photomultiplier tube. The IceCube array consists of 86 
strings, each instrumented with 60 DOMs, which are located at depths 
from $1.45\,\mathrm{km}$ to $2.45\,\mathrm{km}$ below the surface. 
The strings are arranged in a hexagonal pattern with an inter-string 
spacing of about $125\,\mathrm{m}$ and a DOM-to-DOM distance along each string of 
$17\,\mathrm{m}$. A more compact sub-array, called 
DeepCore, consisting of eight densely-instrumented strings, has been 
embedded in the center of IceCube in order to lower the energy threshold from about 
$100\,\mathrm{GeV}$ to about $10\,\mathrm{GeV}$~\cite{DeepCore}. The detector 
construction was completed in December 2010, however data were already 
taken with partial configurations~\cite{Abbasi:2008ym}.
The footprint of IceCube in its 79-string configuration (IC79) is shown 
in Figure~\ref{fig:detector}. This is the configuration used in this analysis. 
Due to its unique position at the geographic South Pole, the zenith 
angle in local coordinates is directly related to the declination and 
the right ascension for a given azimuth angle only depends on the time.


\section{Neutrino flux from dark matter annihilation in the Galactic Halo}
\label{sec:theo}

N-body simulations~\cite{Diemand:2007qr, diemand_clumps_2008, springel_aquarius_2008}
predict the mass distribution $\rho_\mathrm{DM}(r)$ in galaxies as 
function of the distance $r$ to the Galactic Center, assuming a spherically symmetric distribution. 
The resulting dark matter halo profile is parameterized by an extension of the 
Hernquist model~\cite{HaloProfile2}
\begin{equation}
	\rho_{\mathrm{DM}} (r) =
		\frac{\rho_0}
			 {\left ( \delta + {r \over r_s} \right )^\gamma \cdot
			  \left [ 1 +
				\left ( {r \over r_s} \right )^\alpha
			  \right ]^{ (\beta  - \gamma)/\alpha } 
			  },
	\label{eq:rho_profile}
\end{equation}
where $(\alpha, \beta, \gamma, \delta)$ are dimensionless parameters. 
$r_s$ is a scaling radius and $\rho_0$ is the normalization density. 
Both have to be determined for each galaxy.

In this paper the halo profile of Navarro, Frenk and White (NFW)~\cite
{NFW, navarro_universal_1997} with $(1,3,1,0)$ is used as baseline. For the Milky Way 
$r_s=16.1^{+17.0}_{-7.8}\,\mathrm{kpc}$ and $\rho(r_s)=0.47^{+0.05}_{-0.06}\,\GeV/\mathrm{cm}^3$ are used~\cite{nesti_dark_2013}. 
A currently 
favored model is the Burkert profile, that was obtained by the 
observation of dark matter dominated dwarf galaxies. The Burkert profile is described by $(2,3,1,1)$~\cite{Burkert}, 
$r_s=9.26^{+5.6}_{-4.2}\,\mathrm{kpc}$ and $\rho(r_s)=0.49^{+0.07}_{-0.09}\,\GeV/\mathrm{cm}^3$~\cite{nesti_dark_2013}. 
While for the central part of the galaxy the models differ by 
orders of magnitude, the outer profiles are rather similar. 

\begin{figure}
    \includegraphics[width=1.\linewidth]{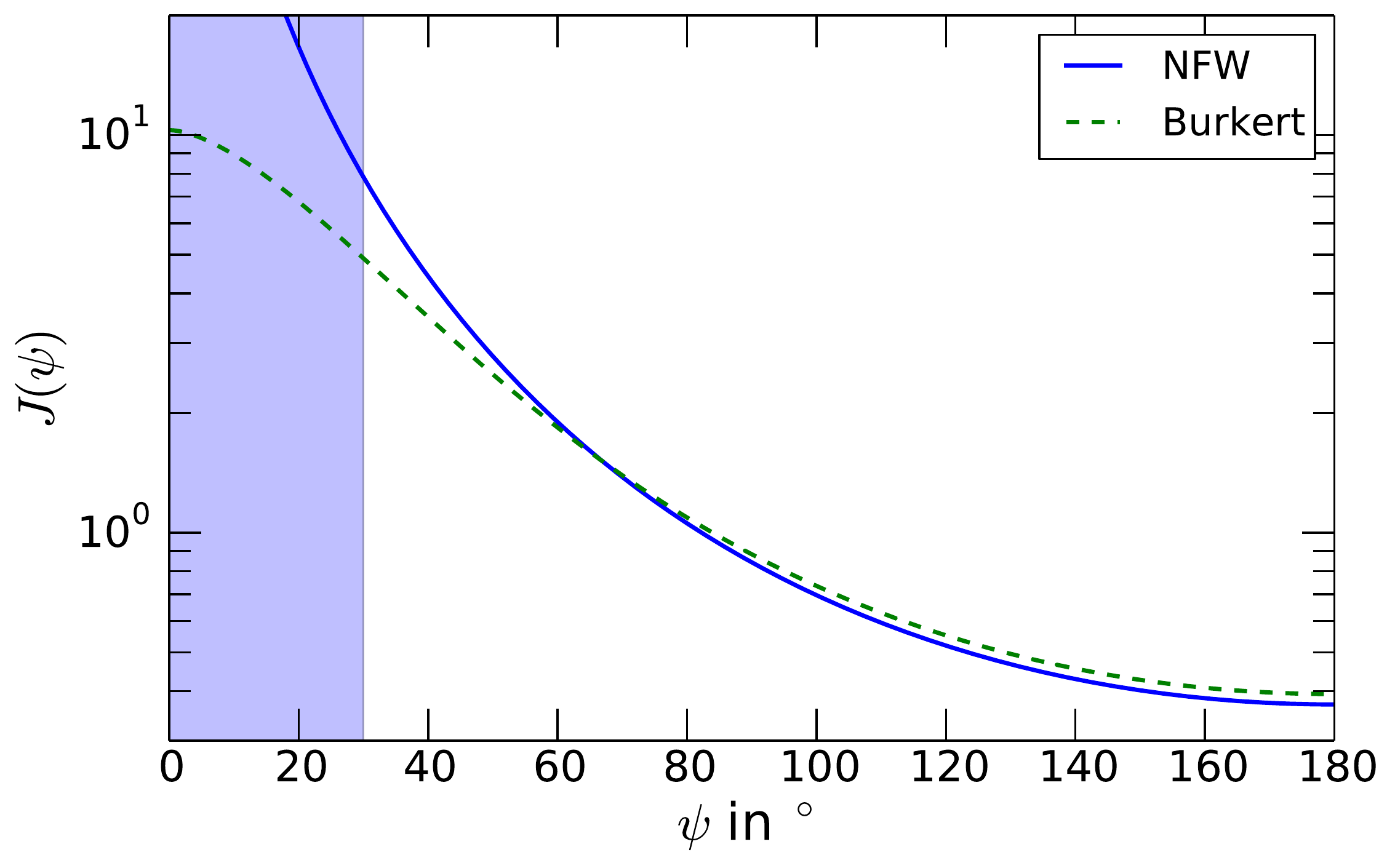} 
	\caption{Line of sight integral $J(\psi)$ as function of the angular 
	distance $\psi$ to the Galactic Center is shown for the different halo profiles 
	used in this analysis. The shaded region corresponds to angular 
	distances to the Galactic Center that lie in the Southern 
	Hemisphere and are not used in this analysis.}
    \label{fig:JofPsi}
\end{figure}

The expected differential neutrino flux $\mathrm{d}\phi_{\nu}/\mathrm{d}E$
at Earth depends on the annihilation rate $\Gamma_A= \sigmav 
\rho(r)^2/2$ along the line of sight $l$, the muon neutrino 
yield per annihilation $\mathrm{d}N_\nu/\mathrm{d}E$, and the self-annihilation cross-section of dark matter
averaged over the velocity distribution $\sigmav$. The flux is given by~\cite{Yuksel}:
\begin{equation}
	\frac{\mathrm{d}\phi_{\nu}}
		 {\mathrm{d}E} =
		\frac{\left\langle \sigma_A v\right\rangle}
			 {2}
		J(\psi)
		\frac{R_{\mathrm{SC}} \rho_{\mathrm{SC}}^2}
			 {4\pi m_{\chi}^2}
		\frac{\mathrm{d}N_\nu}
			 {\mathrm{d}E}\, ,
	\label{eq:flux_formular}
\end{equation}
where $m_\chi$ denotes the mass of the dark matter particle.
$J(\psi)$ is the dimensionless line of sight integral, that 
depends on the angular distance to the Galactic Center, $\psi$, and is defined by~\cite{Yuksel}: 
\begin{equation}
	J(\psi)= \int\limits_0^{d_{\mathrm{max}}} \mathrm{d}l 
		\frac{\rho_{\mathrm{DM}}^2
		\left(\sqrt{R_{\mathrm{SC}}^2-2lR_{\mathrm{SC}} \cos\psi +l^2 }\right)}
		{R_{\mathrm{SC}}\rho_{\mathrm{SC}}^2}
	\,,
	\label{eq:line_of_sight}
\end{equation}
where $\rho_\mathrm{DM}^2$ is evaluated along the line of sight, that is parameterized by $\sqrt{R_{\mathrm{SC}}^2-2lR_{\mathrm{SC}} \cos\psi +l^2 }$ 
and $\rho_{\mathrm{SC}}$ is the local dark matter density at the distance 
$R_{\mathrm{SC}}=8.5\,\mathrm{kpc}$ of the Sun from the Galactic Center~\cite{Yuksel}. $d_\mathrm{max}$ is 
the upper boundary of the integral and is sufficiently larger than the size of the galaxy.
The dimensionless line of sight integral for different halo profiles 
is shown in Figure~\ref{fig:JofPsi}. A large difference for small angles $\psi$
can be seen, while for the outer part a similar factor is expected for all models. 

\begin{figure}
    \includegraphics[width=1.\linewidth]{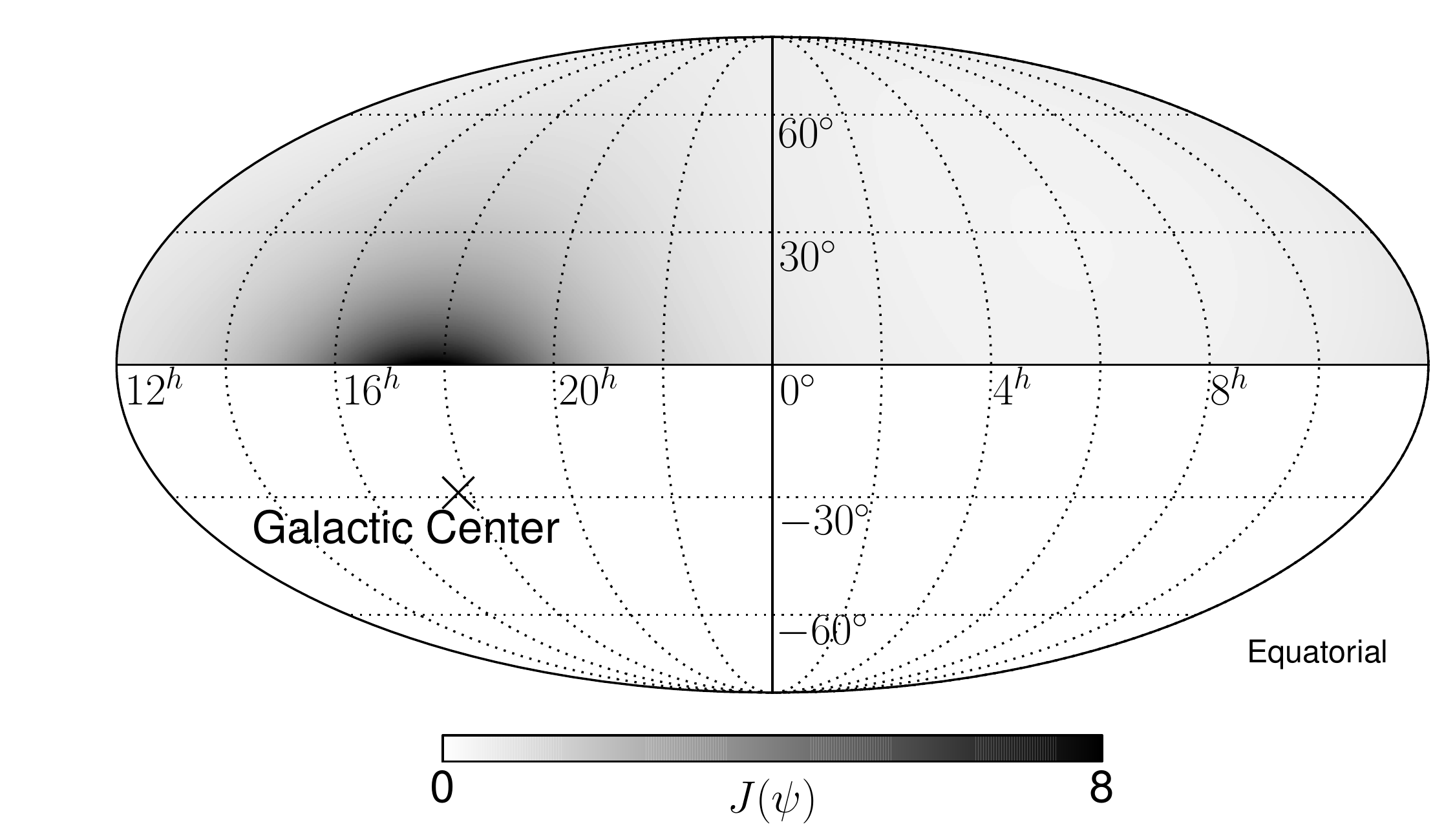} 
	\caption{Dimensionless line-of-sight integral for the NFW 
	profile is shown for the Northern Hemisphere in equatorial 
	coordinates. The anisotropy in the line-of-sight integral causes 
	the anisotropy in the expected flux of neutrinos from self-annihilation of dark matter 
	in the Galactic halo. The position of the Galactic Center is 
	indicated by the cross.}
    \label{fig:JofPsiAnisotropy}
\end{figure}

This analysis searches for an anisotropy in the neutrino arrival directions on the Northern Hemisphere. 
Here we expect a characteristic anisotropy, proportional to 
$J(\psi)$ as shown in Figure~\ref{fig:JofPsiAnisotropy}.

\begin{figure} 
    \includegraphics[width=1.\linewidth]{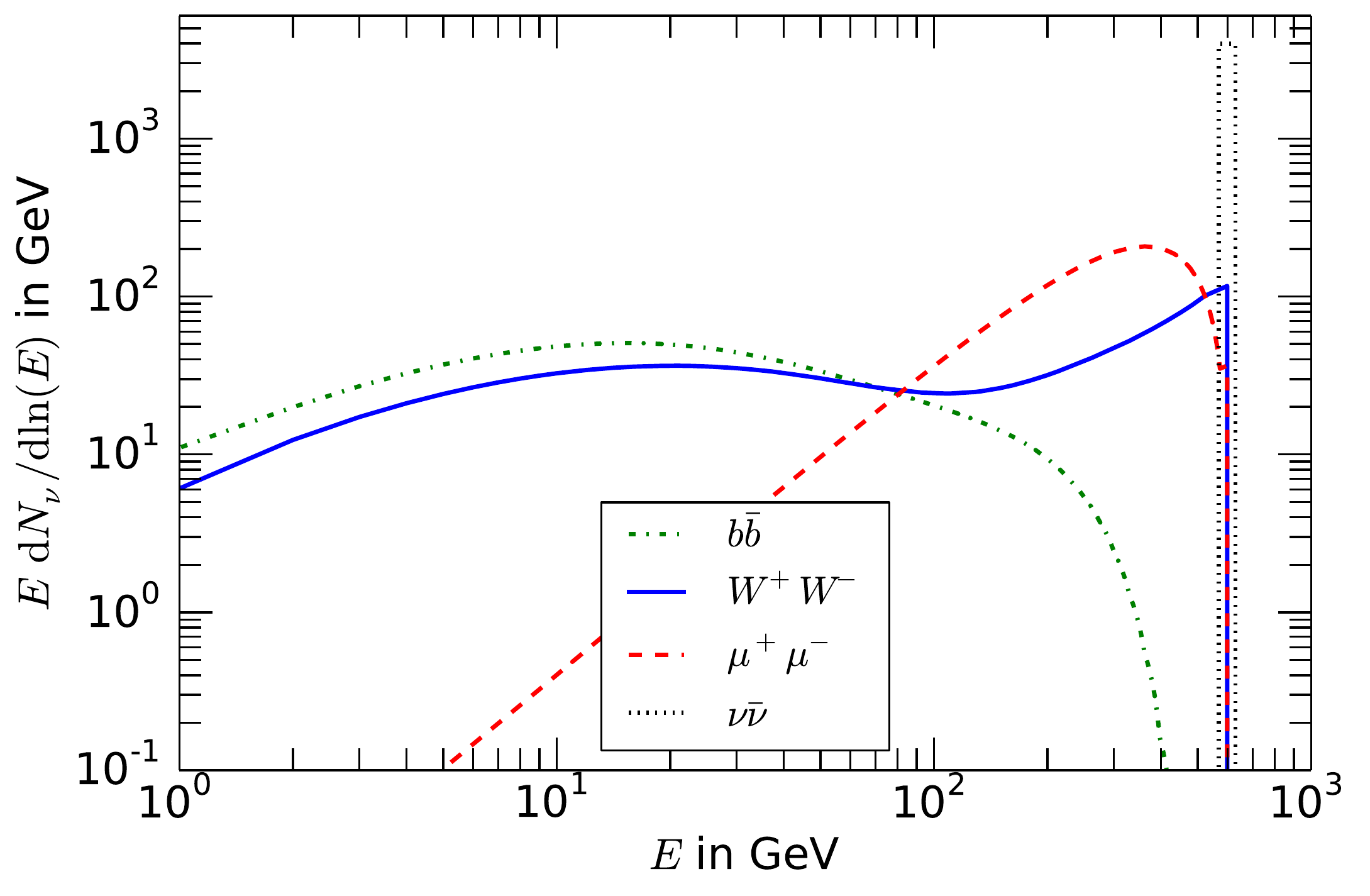} 
	\caption{Muon neutrino multiplicity per annihilation $\mathrm{d}N_\nu/\mathrm{d}\ln(E)$ 
	as function of energy is shown for the investigated benchmark channels and $m_\chi=600\,$GeV. 
	The oscillation probability into muon neutrinos, in the long baseline limit, is included in $\mathrm{d}N_\nu/\mathrm{d}\ln(E)$.
	Beside the neutrino line spectrum, the spectra were calculated 
	with DarkSUSY~\cite{DarkSUSY}.}
    \label{fig:NeutrinoEnergySpectra}
\end{figure}

The neutrino multiplicity per annihilation for the flavors $e,\mu,\tau$,
are obtained with DarkSUSY, which is based on Pythia6~\cite{DarkSUSY,IC22}. 
The muon neutrino multiplicity per annihilation at Earth, $\mathrm{d}N_\nu/\mathrm{d}E$, 
includes the oscillation probability into muon neutrinos in the long baseline limit.
The effective oscillation probability was calculated by numerical averaging of the oscillation probability over a sufficient number of oscillation length using mixing angles and amplitudes from~\cite{PDG}.
Since the nature of the DM particles, as well as the branching ratio for different
annihilation channels, are unknown, a 100\,\% branching ratio to
a few benchmark channels is assumed.
Similar to previous analyses~\cite{aartsen_icecube_2013}, we use the 
annihilation to $b\bar{b}$ as a soft channel, $W^+W^-$ as a medium and 
$\mu^+\mu^-$ as a hard channel.  
Furthermore we investigate direct annihilation to $\nu\bar{\nu}$, which
results in a line spectrum. We assume a 1:1:1 neutrino flavor at source,
and then use the long-baseline approximation as for all other
spectra. This model is implemented as a uniform distribution within $\pm5\,\%$
of $m_\chi$, instead of a Dirac delta-distribution, for computational reasons.
The different muon neutrino multiplicity per annihilation spectra $E^2\mathrm{d}N_\nu/\mathrm{d}E=E\cdot \mathrm{d}N_\nu/\mathrm{d}\ln(E)$ are shown 
in Figure~\ref{fig:NeutrinoEnergySpectra}. 

\section{Data Sample}
\label{sec:sample}

\subsection{Experimental Data}

\begin{figure}
    \includegraphics[width=1.\linewidth]{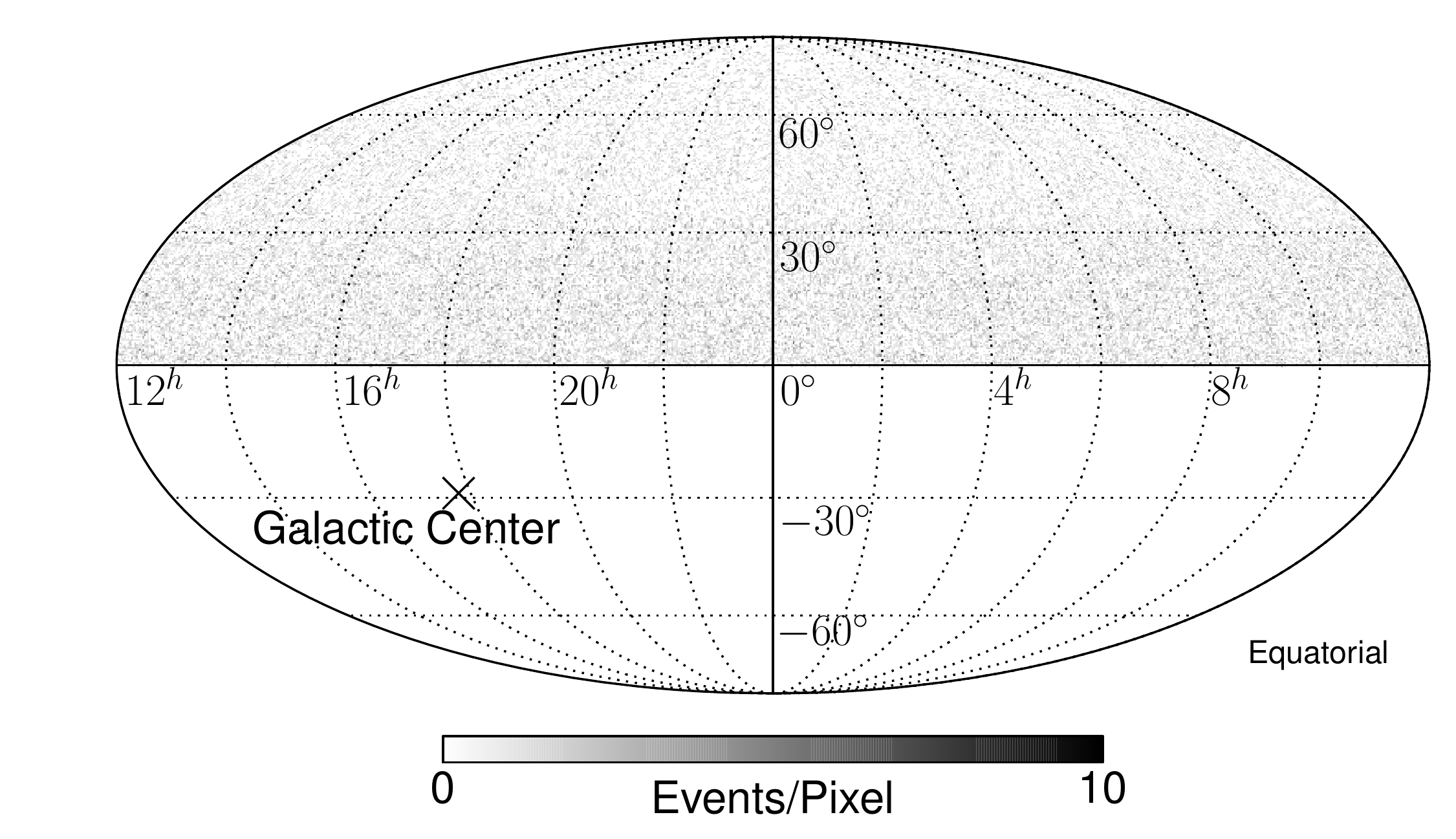} 
	\caption{Sky map of reconstructed neutrino arrival direction of the experimental data 
	sample in equatorial coordinates. The position of the Galactic Center is 
	indicated by the cross.}
    \label{fig:expSkymap}
\end{figure}

Data taken from June 2010 to May 2011, with a total live-time 
$T_{\mathrm{live}}$ of 316 days, are used. Up-going muon events (declination $>$ 0) were 
selected in order to eliminate atmospheric muon background, which becomes dominant at a few degrees above the horizon. By means 
of a mixture of one dimensional cuts on event quality parameters and 
a selection by a Boosted Decision Tree (BDT)~\cite{BDT} the 
contamination of misreconstructed 
atmospheric muons that mimic up-going neutrinos was reduced 
to \mbox{$<3\,\%$~\cite{PSSample2}}. The 
detailed selection is described in~\cite{PSSample} as ``Sample B'' 
for IC79.
After the rejection of atmospheric muons, the sample 
consists of 57281 up-going muon events from the Northern Hemisphere, 
mostly atmospheric muon neutrinos, which are background for the 
search of neutrinos from self-annihilating dark matter in the Galactic halo. 
Unlike signal, the integrated atmospheric neutrino flux is nearly 
constant with right ascension~\cite{IC22}. The reconstructed arrival 
directions of all events in the final sample are shown in Figure~\ref 
{fig:expSkymap}. From full detector simulation it was found that 90\,\%
of the events have a neutrino energy in the range from about $100\,\GeV$ 
to about $10\,\TeV$, with a median of $613\,\GeV$. The median angular 
resolution is $<1^\circ$ for energies above $100\,\GeV$~\cite
{PSSample}. Further details on the sample properties can be found 
in~\cite{PSSample}.

\subsection{Pseudo-Experiments}\label{sec:simulation}

The sensitivity of this analysis has been estimated and optimized by 
pseudo experiments with simulated sky maps of neutrino arrival 
directions. These sky maps contain background from atmospheric 
neutrinos and misreconstructed atmospheric muons and signal from dark matter annihilation.

Signal events are generated at a rate proportional to the line-of-sight integral. 
Furthermore, the arrival direction is smeared  according to the 
angular resolution~\cite{IC22}, which was determined with the
full detector simulation. Moreover, the acceptance of each event is randomized according to the
declination-dependent effective area. It is assumed that the acceptance is constant in RA, 
due to IceCube's special position at the South Pole and the daily rotation 
of the Earth and the almost continuous operation of the detector, which results in a livetime of 91\,\% at final selection level.

The background generation is done by scrambling experimental data. 
Here, the declination of the experimental events are kept and the RA
is uniformly randomized. The 
rescrambling of experimental data to generate background is justified by a negligible
signal contamination in the experimental data. By this technique the background estimation is not 
affected by systematic uncertainties from Monte-Carlo simulation.

The number of signal events in a sky map is fixed to $N_\mathrm{sig}$.
The total number of events in a sky map $N_\nu$ is fixed to the total 
number of events in the experimental sample, so that the sky maps are filled 
up with $N_\nu-N_\mathrm{sig}$ background events.


\section{Method}\label{sec:method}

\subsection{Multipole Expansion of Sky Maps}

The sky maps of reconstructed neutrino arrival directions are expanded by spherical harmonics $Y_\ell^m$~\cite{SPH}. 
Spherical harmonics are given by
\begin{equation}
	Y_\ell^m(\theta,\phi) =
		\sqrt{
			\frac{\left(2\ell+1\right)\left(\ell-m\right)!}
				 {4\pi\left(\ell+m\right)!}
		}
		P_\ell^m\left(\cos\left(\frac{\pi}{2}-\theta\right)\right)
		\mathrm{e}^{\mathrm{i} m\phi}
	\, ,
	\label{eq:spherical_harmonics}
\end{equation}
where $\theta$ is the declination and $\phi$ the RA. $\ell,m$ are 
integer numbers with $0\leq\ell$ and $-\ell\leq m\leq\ell$. 
$P_\ell^m$ are the associated Legendre polynomials. 
Because spherical harmonics are a complete set of orthonormal 
functions, one can expand all square-integrable functions 
$f(\theta,\phi)$ on a full sphere $\Omega$ into spherical harmonics.  
The expansion is given by 
\begin{equation}
	f(\theta,\phi) = 
		\sum_{\ell} \sum_{m=-\ell}^{m=\ell} 
			a_\ell^m \cdot Y_\ell^m(\theta,\phi)
	\label{eq:expantion}
\end{equation}
with expansion coefficients $a_\ell^m$. Here, $\ell$ is the order of the 
expansion and corresponds to an angular scale of approximately 
$180^\circ/\ell$, while $m$ corresponds to the orientation of the 
spherical harmonic. The expansion coefficients are given by
\begin{equation}
	a_\ell^m =
		\int_\Omega \mathrm{d}\Omega\, 
			f(\theta,\phi)\, Y_\ell^{m*}(\theta,\phi)
	\, .
	\label{eq:expand_coef}
\end{equation}
The sky map of reconstructed arrival directions is represented by
\begin{equation}
	f(\theta, \phi) =
		\sum_{i=1}^{N_{\nu}}\, 
			\delta^\mathrm{D}(\cos(\theta)-\cos(\theta_i))
			\cdot\delta^\mathrm{D}(\phi-\phi_i)
	\, ,
	\label{eq:skymap}
\end{equation} where $(\theta_i,\phi_i)$ are reconstructed 
coordinates of event $i$ in equatorial coordinates. $N_\nu$ is the 
total number of events in the data sample and $\delta^\mathrm{D}$ is 
the Dirac-delta-distribution. Since the median angular resolution 
of the events ($<1^\circ$) is much smaller than the anisotropy to 
search for, the usage of Dirac-delta-distributions is justified. 

Coefficients with negative $m$ do not provide additional 
information, because the sky map is described by a real function, 
leading to $|a_\ell^m| = |a_\ell^{-m}|$, and a fixed relation between 
$\mathrm{arg}(a_\ell^m)$ and $\mathrm{arg}(a_\ell^{-m})$~\cite{SPH}.

The multipole expansion is linear and the expansion 
coefficients for signal and background follow the superposition principle. This can be 
seen from Equation~\ref{eq:expand_coef}, if one uses $f(\theta,\phi) 
= s\cdot f_\mathrm{sig}(\theta,\phi) + (1-s)\cdot 
f_\mathrm{bgd}(\theta,\phi)$, where $f_\mathrm{sig}(\theta,\phi)$ is 
the sky map for pure signal, $f_\mathrm{bgd}(\theta,\phi)$ is the 
sky map for pure background, and $s$ is the relative signal strength. 

In practice the expansion is stopped at some large $\ell_\mathrm{max}$. 
Information on structures of an angular scale smaller $180^\circ/\ell$ will be lost. Hence, the value of 
$\ell_\mathrm{max}$ should be sufficiently large to include all 
angular scales of interest.

\subsection{Application of Multipole-Expansion to Pseudo-Experiments}

\begin{figure}
    \includegraphics[width=1.\linewidth]{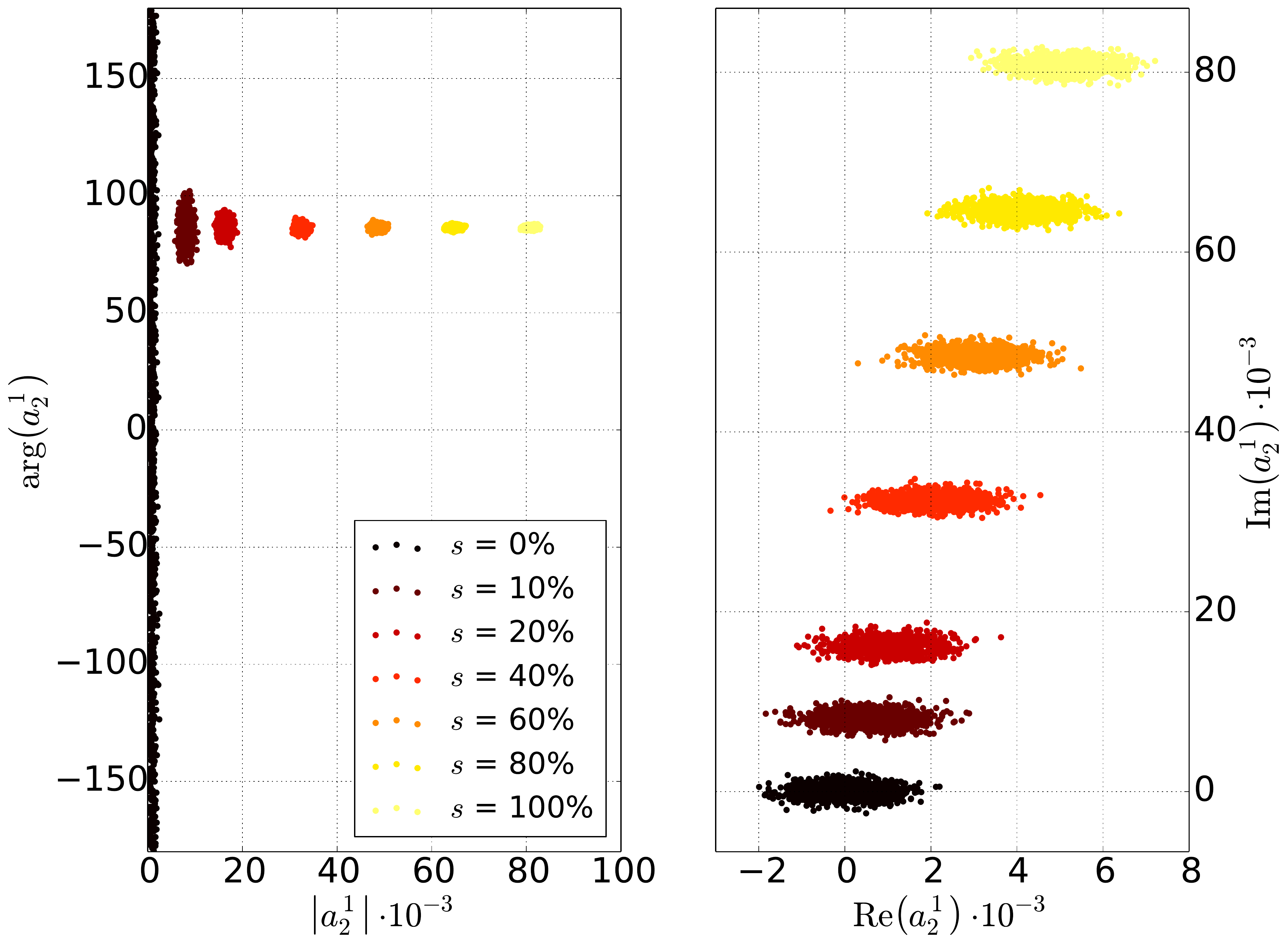}  
	\caption{The expansion coefficient $a_2^1$ for large signal strength $s$ in the Euler 
	representation (left panel) and in the complex plane (right panel). For each 
	signal strength $s$ 1000 pseudo-experiments were 
	generated and the expansion coefficient $a_2^1$ calculated.}
    \label{fig:example_coefB}
\end{figure} 
\begin{figure}
    \includegraphics[width=1.\linewidth]{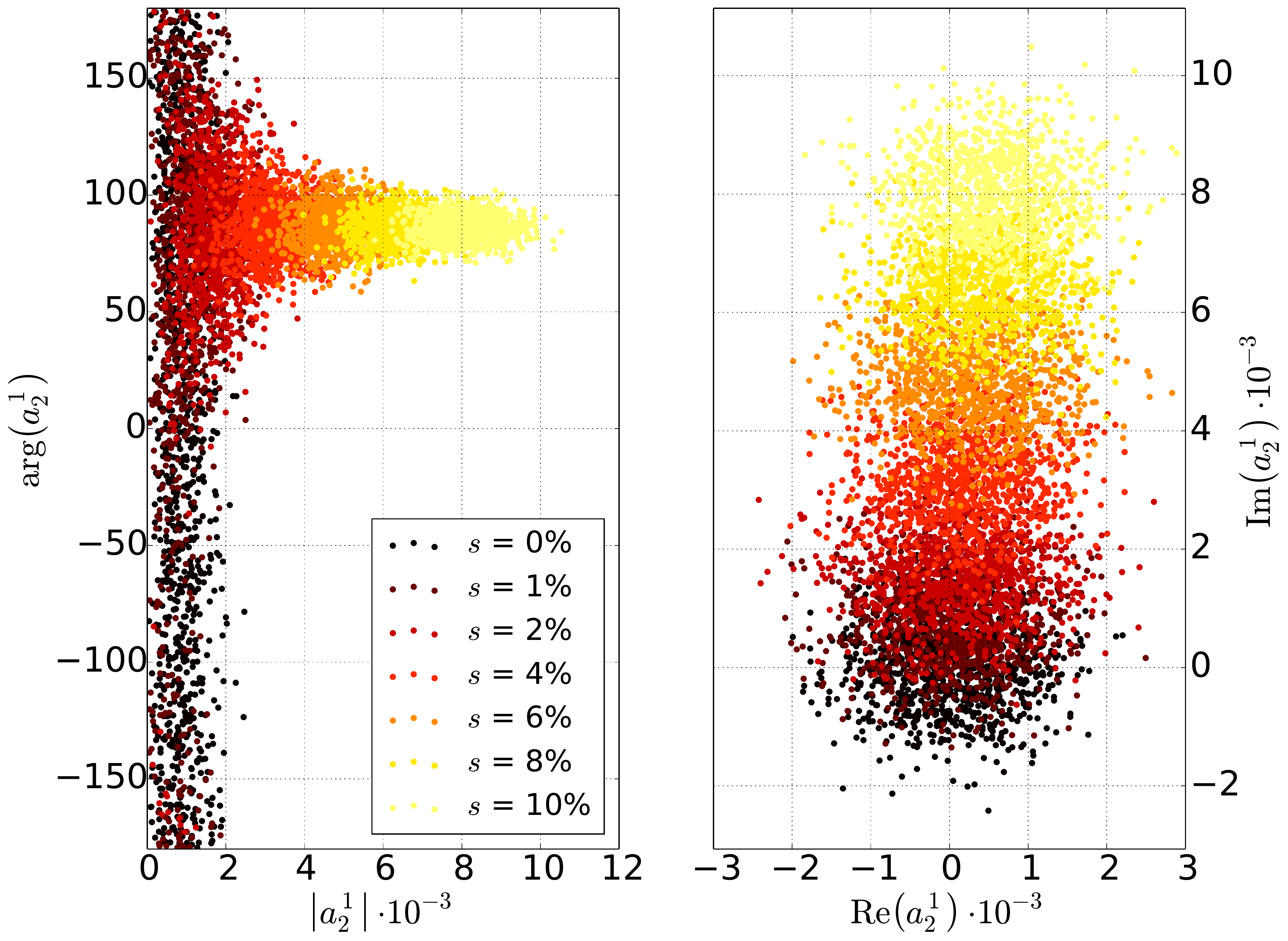}  
	\caption{The expansion coefficient $a_2^1$ for small signal strength $s$ in the Euler 
	representation (left panel) and in the complex plane (right panel). For each 
	signal strength $s$ 1000 pseudo-experiments were 
	generated and the expansion coefficient $a_2^1$ calculated.}
    \label{fig:example_coefA}
\end{figure} 

For this analysis the calculation of the expansion coefficients is done 
with the software package HealPix~\cite{HealPixURL,HealPix}. 

Figure~\ref{fig:example_coefB} and Figure~\ref{fig:example_coefA} show the expansion coefficient 
corresponding to $Y_\ell^m$ with $\ell=2$ and $m=1$ for a
signal, as described in Section~\ref{sec:theo}, and a uniform  
distributed background in RA, as described in Section~\ref{sec:sample}.

For different signal strength $s=N_\mathrm{sig}/N_\nu$, 1000 pseudo-experiments were performed and 
$a_2^1$ was calculated. 
For pure background sky maps ($s=0\,\%$) with no preferred direction in RA (uniform) the 
expansion coefficient shows no preferred phase and is almost normal distributed 
around the origin of the complex plane. For pure signal ($s=100\,\%$) 
there is a clear separation from the origin. Also, a clear preferred phase 
can also be observed, which corresponds to the orientation of the expected anisotropy in the sky. This phase is the same as the preferred phase for 
the sky maps with partial signal ($0\,\%<s<100\,\%$). 
Furthermore, a
linear dependency between the signal strength $s$ and 
the mean power $\langle |a_2^1| \rangle$ can be seen.

\begin{figure*}
    \includegraphics[width=0.5\linewidth]{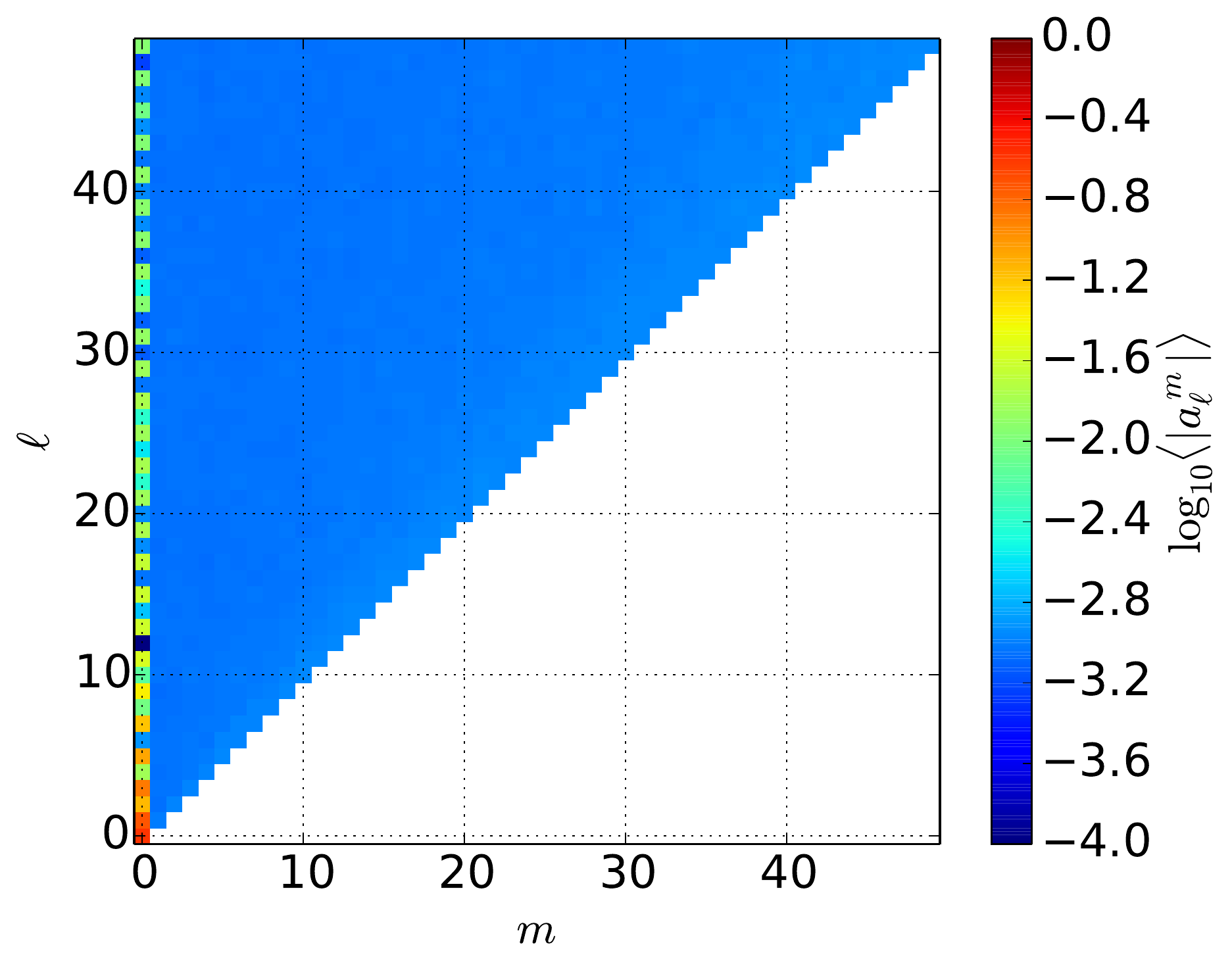} 
    \includegraphics[width=0.5\linewidth]{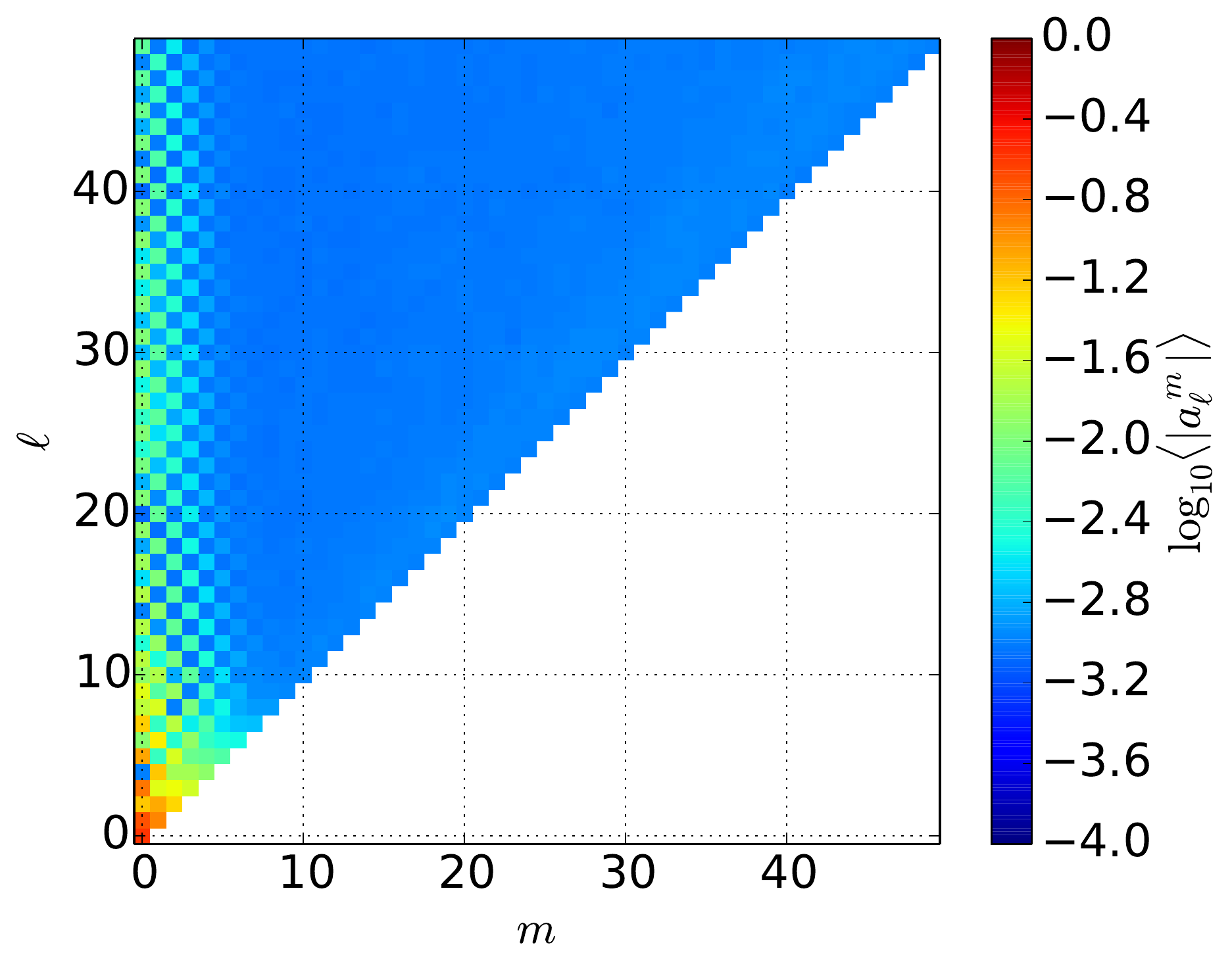} 
	\caption{Mean logarithm of the absolute value for all 
	expansion coefficients in the $\ell$-$m$-plane up to $\ell,m=50$. 
	The mean absolute value was obtained from 1000 pure background/signal
	 sky maps (left/right).}
    \label{fig:alm_abs_bgd}\label{fig:alm_abs_sig}
\end{figure*} 

In practice the number of events $N_\nu$ in the map is limited. 
Therefore, the value of $a_\ell^m$ has a statistical error, which depends  
on the total number of events in the sky map $N_\nu$ and 
weakly on the signal fraction $s$. For the value $N_\nu > 57000$ of 
this analysis the error can be well approximated as Gaussian.

An overview of the logarithm of the absolute value of all expansion 
coefficients with $0\leq \ell, m \leq 50$ is shown in Figure~\ref{fig:alm_abs_bgd} 
for pure background (left panel), and pure signal (right panel). 
For the pure background case 
most of the power is contained in the 
coefficients $a_\ell^0$, related to the pure zenith structure. 
Because the background was assumed to be isotropic in RA all coefficients with  $m\neq0$ are at noise level. 
The statistical noise level in the map is of the order of $10^{-3} - 10^{-4}$
and corresponds to the width of the distribution, as shown in Figure~\ref{fig:example_coefB} and Figure~\ref{fig:example_coefA}. 

From equation~(\ref{eq:spherical_harmonics}) one 
can see that spherical harmonics with $m=0$ are independent of RA 
and thus purely depend on declination. The $a_\ell^0$ coefficients have an 
absolute value larger than the noise level that means they contain power. 
These coefficients describe the full declination structure, that is mainly 
influenced by the declination-dependent acceptance and the declination-dependent variation of the atmospheric neutrino flux. 
Furthermore it was found that there is no preferred phase in any 
coefficient for background.

For pure signal (Figure~\ref{fig:alm_abs_bgd}, right) there is also 
power in coefficients with $m\neq0$, resulting from the 
characteristic anisotropy of the signal. It was found that all 
coefficients that have a power larger than the noise level also have 
a preferred phase. The characteristic checkered pattern in the 
coefficients results from the observation of just one hemisphere $\theta>0$, 
leading to a suppression of coefficients with even $\ell+m$, that 
correspond to a symmetric spherical harmonic with respect to the 
equator.

From Figure~\ref{fig:alm_abs_bgd} (right panel) it becomes obvious that 
coefficients with small $\ell$ and $m$ carry most power. This is due to the 
large-scale anisotropy of the line-of-sight integral (see Figure~\ref{fig:JofPsiAnisotropy}). In analogy to the 
relation of $\ell$ and the characteristic angular scale of the 
structure, $m$ is related to the characteristic angular scale in RA, 
thus small $m$ represent large structures in RA and large $m$ represent small structures in RA.

\subsection{Test Statistic}

\begin{figure}
    \includegraphics[width=1.\linewidth]{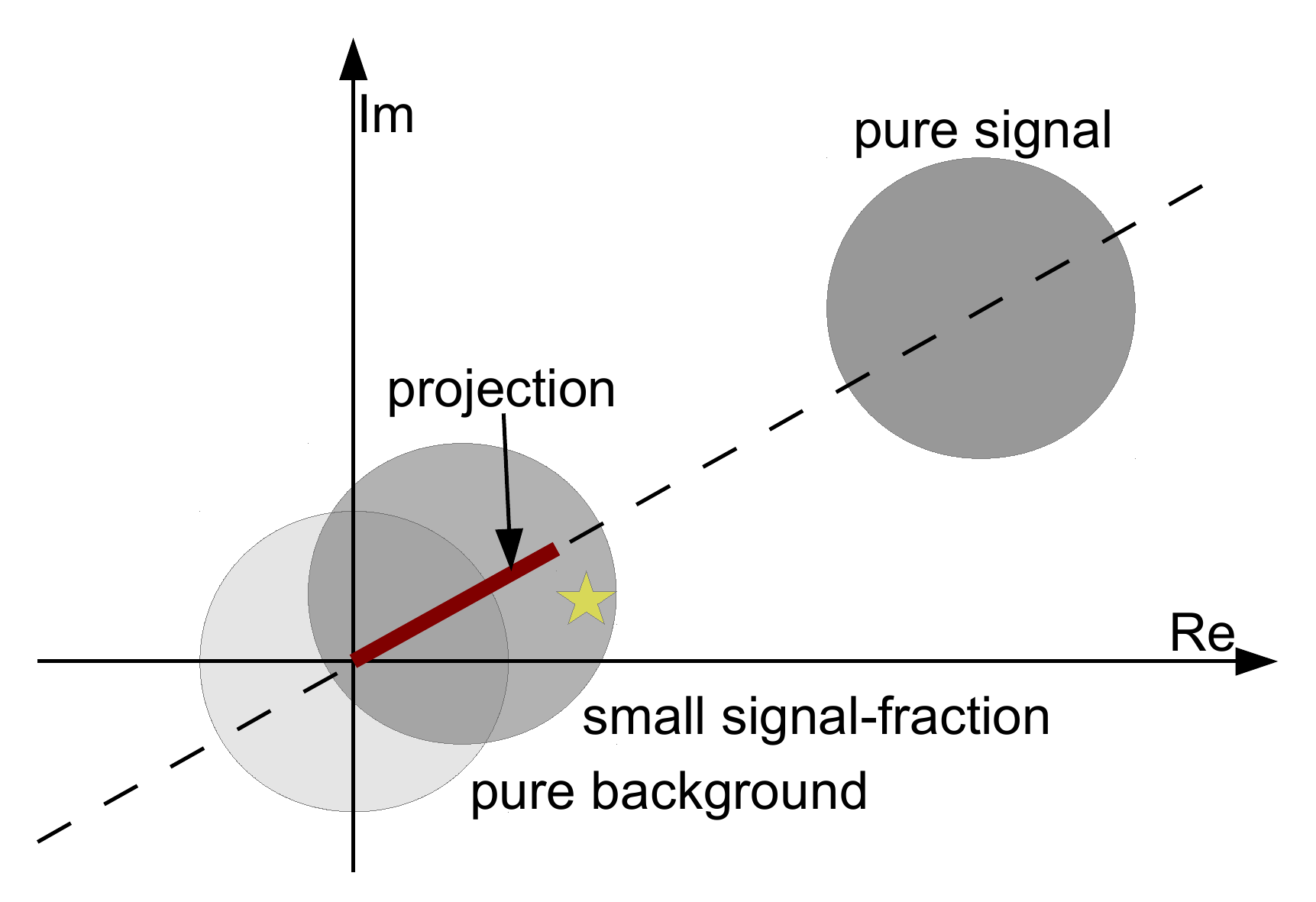}
	\caption{Sketch to illustrate the projection of complex expansion 
    coefficients, on the axis corresponding to the preferred direction. 
    The large gray circles represent the central part of the Gaussian in the complex plane
    in a large ensemble limit for different signal strength. 
    In contrast to that distributions the star corresponds to one 
    specific value of the expansion coefficient for one measured sky map. 
    The projected value of this specific expansion coefficient is given 
    by the length of the thick red line.}
    \label{fig:proj_sketch}
\end{figure} 

The test statistic ($TS$) to separate signal from background combines the 
phase information and the power of a complex coefficient 
into one value. A projection of the complex coefficient onto the axis, 
corresponding to the preferred phase, is introduced~\cite{Rene}. This projection is 
illustrated in Figure~\ref{fig:proj_sketch} and is given by
\begin{equation}
	\mathcal{A}_{\ell}^m =
		\left\|a_\ell^m\right\|
		\cos\left(
			\arg\left(a_\ell^m\right)
			- \left\langle\arg\left(a_{\ell,\mathrm{sig}}^m\right)\right\rangle
		\right)
    \, ,
	\label{eq:projection}
\end{equation}
where $\arg\left(a_\ell^m\right)$ is the argument of $a_\ell^m$ and 
$\langle\arg(a_{\ell,\mathrm{sig}}^m)\rangle$ is the mean expected phase of 
the $a_\ell^m$ of pure signal pseudo-experiments. 

This projected expansion coefficient has the following advantages. 
First $\mathcal{A}_{\ell}^m$ is proportional to the power of 
the expansion coefficient. Second, the most sensitive direction is 
the axis of the preferred phase, and the value of the projection 
gets smaller, the more the phase differs. This results in negative 
values for $\mathcal{A}_{\ell}^m$, if the phase differs more than 
$\pi/2$, indicating that the anisotropy is in the opposite 
direction of the expectation. 

Using these projected expansion coefficients the $TS$ is 
defined as 
\begin{equation}
	TS = \frac{1}{\sum w_\ell^m}
		 \sum_{\ell=1}^{\ell_{\mathrm{max}}} \sum_{m=1}^{\ell} 
			\mathrm{sig}\left(\mathcal{A}_{\ell}^m\right)
			w_\ell^m
			\left(
				\frac{\mathcal{A}_{\ell}^m - \left\langle \mathcal{A}_{\ell,\mathrm{bgd}}^m
				\right\rangle} {\sigma\left(\mathcal{A}_{\ell,\mathrm{bgd}}^m\right)}
			\right)^2
    \,
	\label{eq:test_statistic}
\end{equation}
where $\mathrm{sig}(x)$ gives the sign of $x$ and
$\langle \mathcal{A}_{\ell,\mathrm{bgd}}^m\rangle$ and 
$\sigma(\mathcal{A}_{\ell,\mathrm{bgd}}^m)$ are the mean and standard 
deviation of an ensemble of $\mathcal{A}_\ell^m$ estimated from pseudo-experiments 
of pure background~\cite{Rene}. $w_\ell^m$ are individual weights for each 
coefficient. The definition of the test statistic is motivated by a 
weighted $\chi^2$-function. The weights are chosen with respect to the 
separation power of the different coefficients and are defined below. 
Because the sign of the deviation is lost in the 
squared term, the sign is included as an extra factor.  
Coefficients with no power, especially in the background case, have 
randomly positive or negative sign. In average they add up to zero, 
however for signal always positive values contribute to the sum. 

The weights are given by
\begin{equation}
	w_\ell^m =
	\left|
		\frac{\left\langle \mathcal{A}_{\ell,\mathrm{sig}}^m \right\rangle- \left\langle
		\mathcal{A}_{\ell,\mathrm{bgd}}^m \right\rangle}
		{\sigma\left(\mathcal{A}_{\ell,\mathrm{bgd}}^m\right)}
	\right|
    \,,
	\label{eq:weight}
\end{equation}
where $\langle \mathcal{A}_{\ell,\mathrm{sig}}^m \rangle$ is the 
expected projected expansion coefficient for pure signal, that can be 
calculated by averaging over the $\mathcal{A}_\ell^m$ of an ensemble of 
pseudo-experiments for pure signal. 
Because $\mathcal{A}_\ell^m$ is proportional to $a_\ell^m$ 
and, thus, to the signal strength, a smaller signal expectation would just 
lead to a different normalization of the weight, which is absorbed in 
the factor $1/\sum w_\ell^m$. Therefore the relative strength of 
coefficients in this test statistic does not depend on the signal strength $s$. 
The weights represent the power to separate signal from 
background for each coefficient. Insensitive expansion coefficients get assigned a small weight and do not contribute.

\subsection{Test statistic application to the search for dark matter}

\begin{figure}
    \includegraphics[width=1.\linewidth]{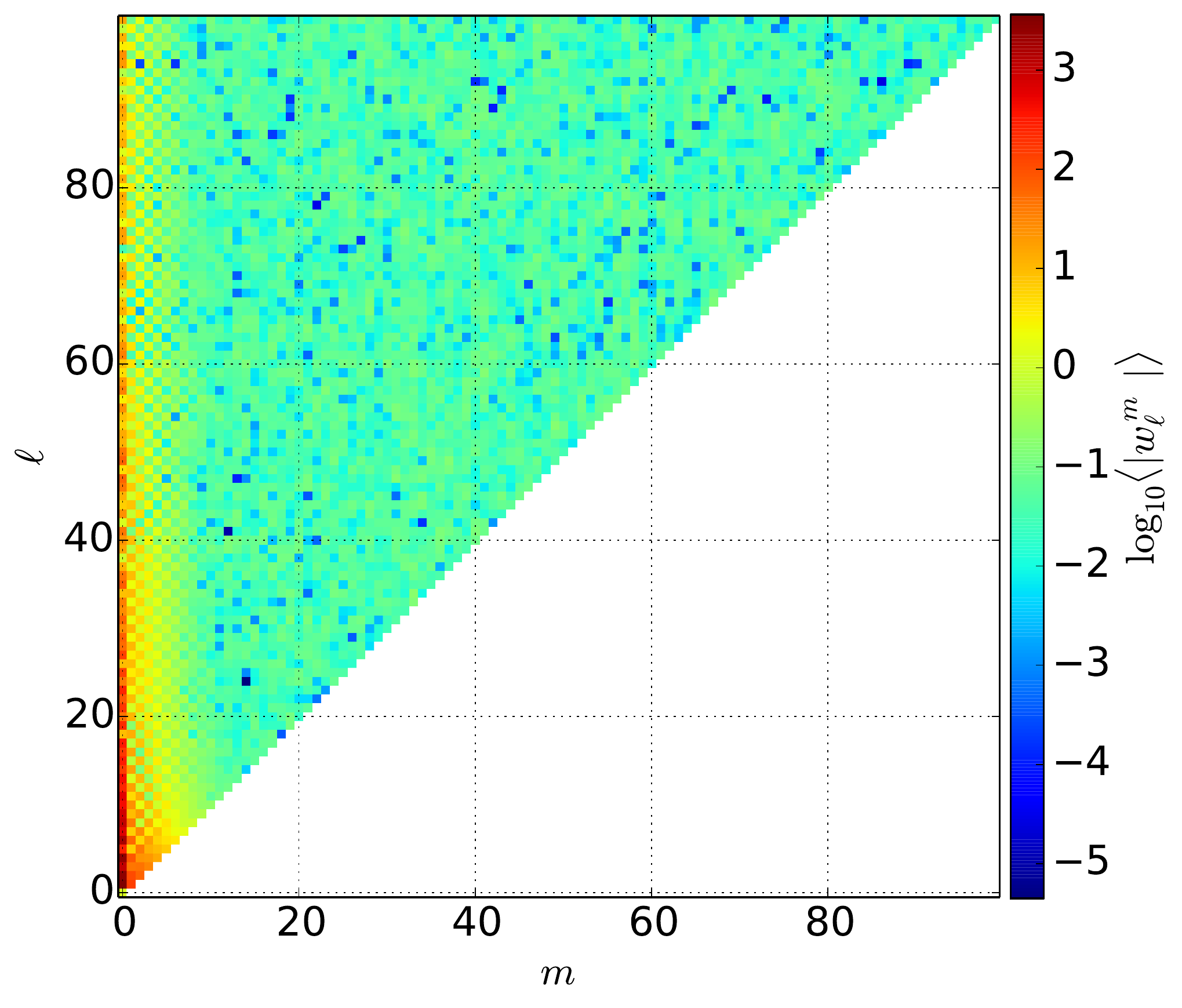} 
    \caption{The logarithm of the weight (as defined in 
    Equation~\ref{eq:weight}) for all coefficients in the 
    $\ell-m$-plane up to $\ell,m=100$. In the calculation the NFW profile was used.}
    \label{fig:weights}
\end{figure}

To determine $\langle\arg(a_{\ell,\mathrm{sig}}^m)\rangle$, $w_\ell^m$, 
$\langle \mathcal{A}_{\ell,\mathrm{bgd}}^m\rangle$ and 
$\sigma(\mathcal{A}_{\ell,\mathrm{bgd}}^m)$, 1000 pseudo-experiments 
were used in each case. The weights for an NFW profile for all coefficients with 
$\ell,m\leq100$ are shown in Figure~\ref{fig:weights}. 

The weights range over orders of 
magnitude and are proportional to the values shown in Figure~\ref{fig:alm_abs_sig} (right panel) 
reflecting the separation power of the coefficients.

For IceCube, the coefficients with $m=0$ contain the declination dependence
and, due to the detector location at the geographic South Pole, this translates
directly into the zenith dependence of the detector acceptance. In order
to avoid introducing a zenith-dependent systematic uncertainty, the
coefficients with $m=0$ are omitted in this analysis and are not
included in equation~(\ref{eq:test_statistic}). Since spherical harmonics
are orthonormal functions, no additional systematic is introduced by
this choice. Possible systematic uncertainties in azimuth average out due
to the daily rotation of Earth, and thus the detector.

Because the anisotropy introduced by the flux from dark matter annihilation in 
the halo is a large scale anisotropy a maximal expansion order of 
$\ell_\mathrm{max}=100$ was chosen. 
In general the coefficients become less sensitive with lager $\ell$. Due 
to this generic suppression of insensitive coefficients in the 
test statistic $\ell_\mathrm{max}$ does not need to be optimized.

Since the differences in weights for different halo profiles are found to be  
small, which is a result of the similar shapes of the outer halo predicted by the different models (see Fig.~\ref{fig:JofPsi}) , weights from NFW profiles are used for all model tests to avoid trial factors.
Differences with respect to the halo profiles will be discussed below.

\begin{figure}
    \includegraphics[width=1.\linewidth]{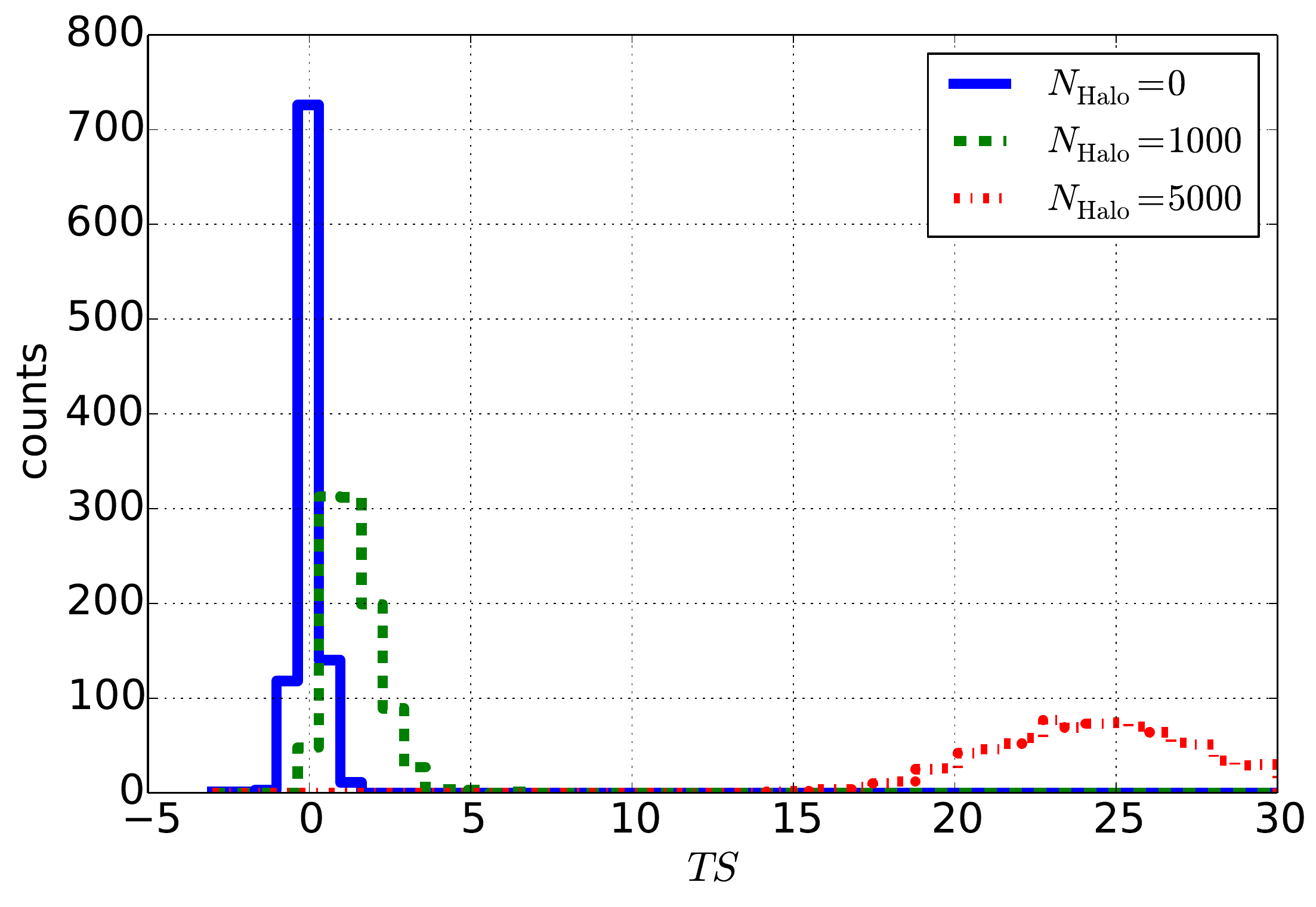} 
    \caption{Test statistic $TS$ for pure background simulation (solid) and
    simulations with small signal contributions assuming a NFW profile. $N_{\mathrm{sig}}$ is the number of 
    simulated neutrino arrival directions from dark matter annihilation in the Galactic halo.}
    \label{fig:TestStatistics}
\end{figure}

Figure~\ref{fig:TestStatistics} shows the 
resulting test statistic for pseudo-experiments of pure background 
($N_\mathrm{sig}=0$, $s=0\,\%$) and pseudo-experiments with signal 
contribution of $N_\mathrm{sig}=1000,5000$ 
(signal strength $s=1.7\,\%,8.7\,\%$) assuming a NFW profile.

\subsection{Generalization of the method}

In the previous sections the assumed signal was the characteristic 
anisotropy of the flux from dark matter annihilation. However, the method described here can be generalized to any 
other anisotropy of preferred direction.

If there is no preferred direction in the signal expectation, i.e. a 
characteristic event correlation structure which
is distributed isotropically on the sky, the phase 
is also random in the signal case.
This is the case e.g. in a search for many point-like sources that are too weak to be detected individually, but which lead to a clustering of events on specific angular scales.
Even in these cases it is possible to define 
a test statistic analogously. Here one can use the averaged power on 
a characteristic scale $C_\ell^\mathrm{eff}$, that is given by
\begin{equation}
    C_\ell^\mathrm{eff} = 
        \frac{1}{2\ell} 
        \sum_{{m=-\ell} \atop {m\neq0}}^{\ell}
            \left| a_\ell^m \right|^2
    \,.
    \label{eq:C_ell}
\end{equation}
Note that also here the power coefficients are defined without the $a_\ell^0$ 
coefficients, resulting in coefficients that are not affected 
by systematic uncertainties in the declination acceptance. In the 
test statistic (equation~(\ref{eq:test_statistic})) one has to replace 
all $\mathcal{A}_\ell^m$ by $C_\ell^\mathrm{eff}$ and $w_\ell^m$ by 
$w_\ell$ and remove the sum over $m$. Furthermore the 
$\mathrm{sig}(\mathcal{A}_\ell^m)$-term now has to be written as 
$\mathrm{sig}(C_\ell^\mathrm{eff}-C_{\ell,\mathrm{bgd}}^\mathrm{eff})$.
The weight $w_\ell$ can be defined similar to equation~(\ref{eq:weight})
by replacing $\mathcal{A}_\ell^m$ by $C_\ell^\mathrm{eff}$. An example 
where such a test statistic has been used is~\cite{Martin, CorrelationPaper}.


\section{Sensitivity}\label{sec:sensitivity}

\begin{figure}
    \includegraphics[width=1.\linewidth]{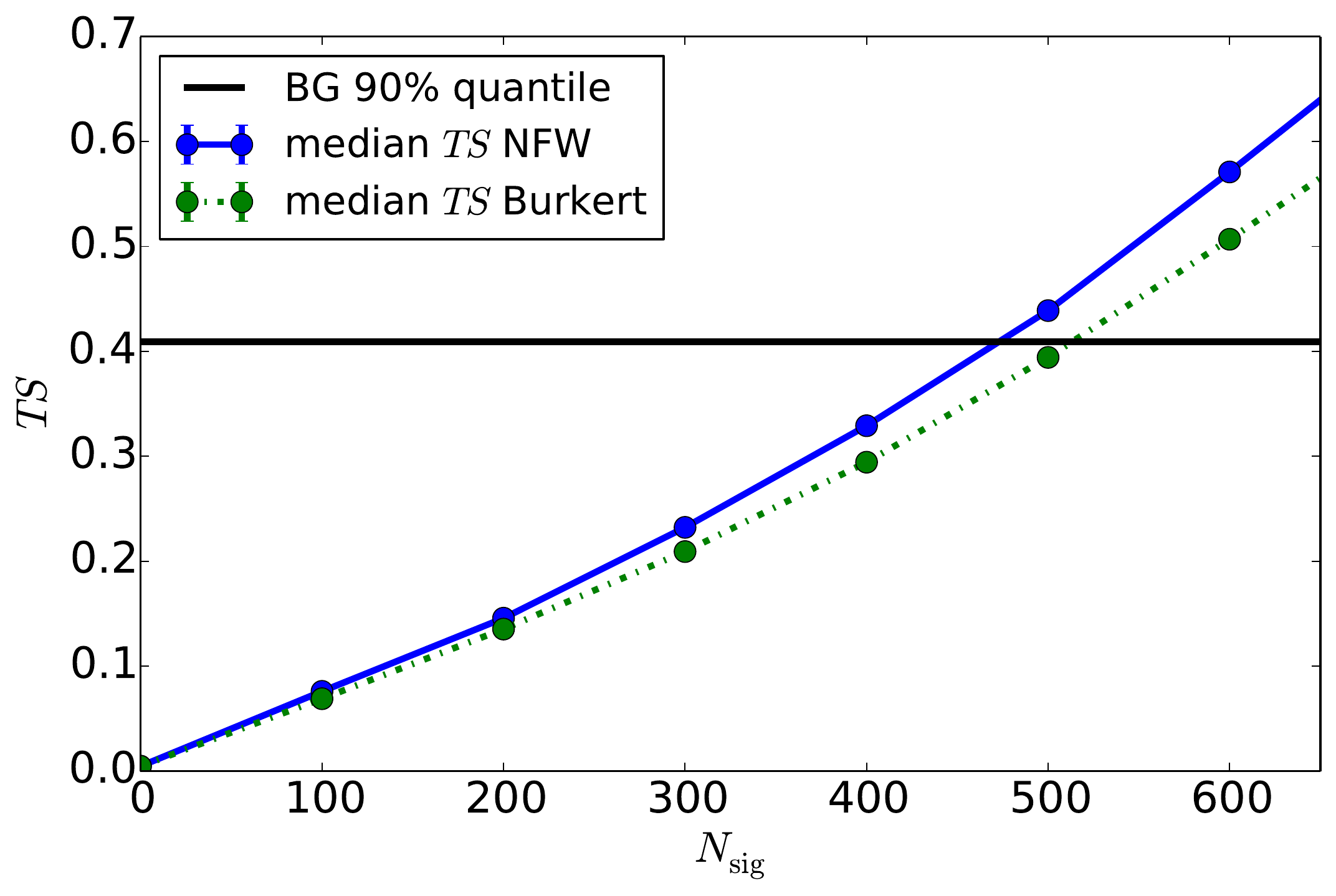} 
    \caption{The median of the test statistic, $TS$, 
    distribution as a function of the signal strength $N_{90}$ for 
    different halo profiles. Furthermore the 90\,\%-upper-quantile of 
    the $TS$ background distribution is shown. 
    The statistical errors were computed by binomial statistics and are smaller than
    the size of the markers.}
    \label{fig:N90}
\end{figure}

The median sensitivity at a 90\,\% confidence level (C.L.), $N_{90}$, 
is given by that number of signal events, where 
50\,\% of the signal $TS$ distribution is larger than the 90\,\% upper 
quantile of the background distribution. To estimate this median 
sensitivity for different signal contributions 
25000 pseudo-experiments were generated for different numbers of signal events $N_\mathrm{sig}$. In 
Figure~\ref {fig:N90} the median of each $TS$ distribution is shown 
versus the signal strength $N_\mathrm{sig}$ for the different halo 
profiles. Further, the 90\,\%-quantile of the background distribution is 
shown. The resulting $N_{90}$ for the different halo profiles are shown in 
Table~\ref{tab:N_sens}. It can be seen that differences in the value of 
$N_{90}$ are smaller than 10\,\%.
Note that the $N_{90}$ value does not depend on the overall normalization but only on the different shape of the profiles.

\begin{table}
    \caption{Median sensitivity on the number of signal events  
    at a 90\,\% C.L. $N_{90}$ and the statistical uncertainty for different halo profiles.}
	\label{tab:N_sens}
    \centering
	\begin{tabular}{ l | r }
		\hline\hline
		Halo Profile & $N_{90}$ \\
		\hline
		NFW & $470.8 \pm 2.3$ \\
		Burkert & $511.2 \pm 2.8$ \\
		\hline\hline
	\end{tabular}
\end{table}

\begin{figure}
	\includegraphics[width=1.\linewidth]{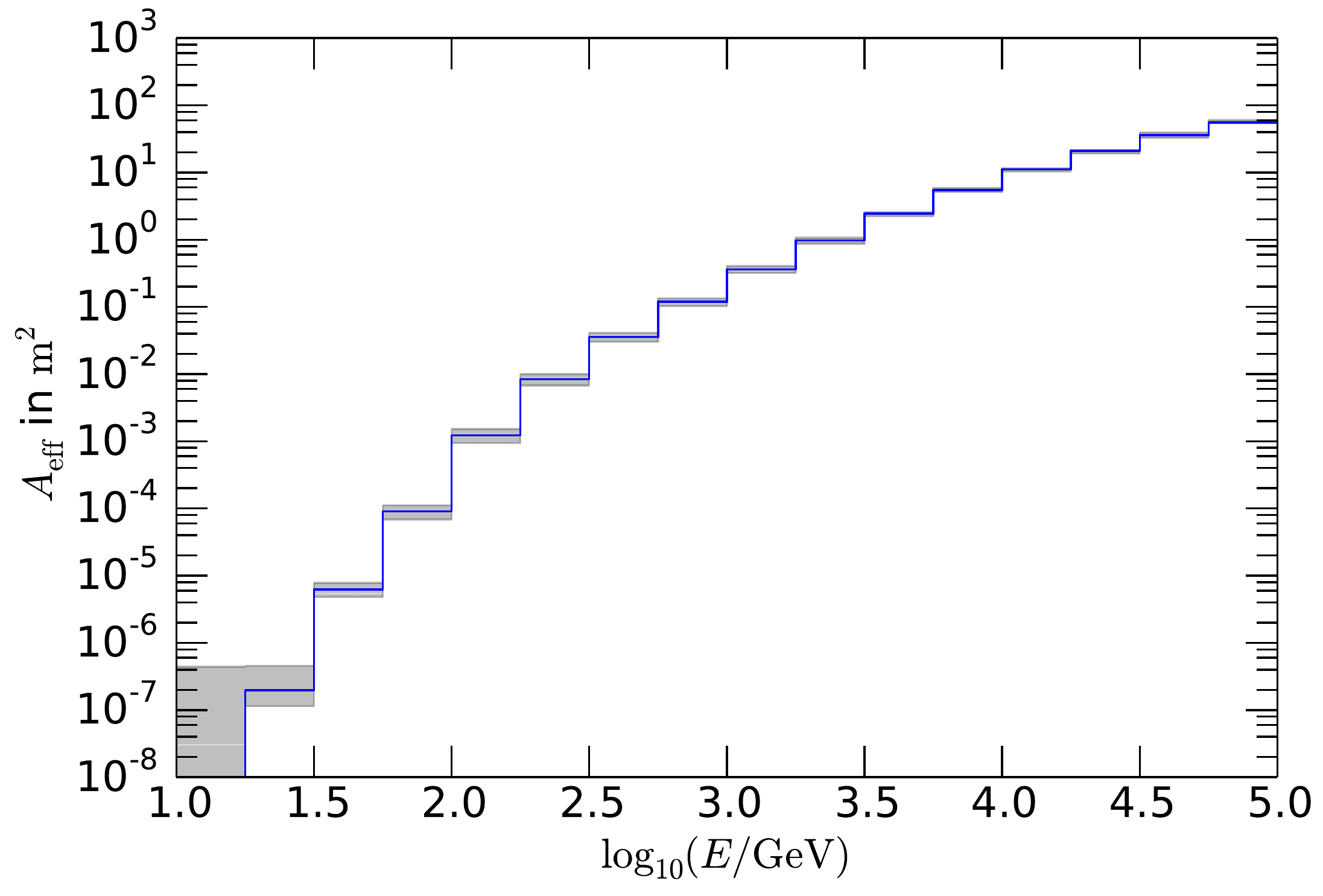} 
	\caption{Effective area as a function of neutrino energy, averaged over
 declination in the Northern Hemisphere. The gray band represents the uncertainties due to systematic uncertainties in the optical detection efficiency and in the ice properties.}
	\label{fig:Aeff}
\end{figure}

The sensitivity on the number of signal events in the data set, 
and thus the flux, can be interpreted in terms of the self-annihilation 
cross-section of the dark matter. Using equation~(\ref{eq:flux_formular}) 
the self-annihilation cross-section is given by
\begin{equation}
	\left\langle\sigma_Av\right\rangle = 
		\frac{8\pi m_\chi^2}{R_{\mathrm{SC}}\rho_{\mathrm{SC}}^2}
		\frac{1}{T_{\mathrm{live}}}
		\frac{1}{\int \int J(\psi) A_{\mathrm{eff}}
		\frac{\mathrm{d}N_{\nu}}{\mathrm{d}E}\mathrm{d}E \mathrm{d}\Omega} N_{90}
    \, .
	\label{eq:sigmav_90}
\end{equation}
Here, $A_\mathrm{eff}$ is the effective area, which is shown for
the chosen data set, averaged over the Northern Hemisphere, 
in Figure~\ref{fig:Aeff}. The resulting sensitivity on the 
self-annihilation cross-section depends on the assumed 
annihilation channel and WIMP mass.

In N-body simulations of structure formation using DM, self-similar substructures are found.
These structures lead to an enhanced annihilation probability, 
because the gain of flux from denser regions is larger than the loss 
in dilute regions. 
The increase of the annihilation rate can be described by a boost factor 
$B(r)$, which modifies the line of sight integral. 
An example of such a boost factor is given 
in~\cite{kamionkowski_galactic_2010} and has been discussed 
in~\cite{IC22}. For this analysis the modification of the 
line-of-sight integral as described in~\cite{IC22} results in a 
median sensitivity, $N_{90}$, which is about 50\,\% worse, 
because the shape of the line-of-sight integral and thus the anisotropy becomes flatter. However, due to the 
larger total expected flux the sensitivity on the 
self-annihilation cross-section is 20\,\% more stringent. 
To be conservative, the results presented here do not take substructures into account.

As a cross-check, the sensitivity on the number of signal events of a cut-and-count based method as described in~\cite{IC22} was determined. 
About 14600 neutrinos are expected in the off-source region, which covers 1.6\,sr. 
This results in a sensitivity to approximately 219 neutrinos when subtracting the number of events in the on-source and off-source regions. 
Taking into account the different solid angles in the denominator of equation~\ref{eq:sigmav_90}, 12\,\% more 
signal events over background are required for the same significance, resulting in a slightly less  sensitive analysis. 


\section{Systematic Uncertainties}\label{sec:systematics}

The relevant systematic uncertainties for the analysis can be categorized into
three groups: 
\begin{itemize}
	\item Systematic uncertainties on the background 
expectation. 
    \item Systematic uncertainties on the signal 
detection efficiency. 
    \item Dependencies on the halo profile. 
\end{itemize}

\begin{table}
	\caption{Systematic uncertainties resulting from pre-existing 
	anisotropies in the experimental sky map. $\sigmav_\mathrm{base}$ 
	denotes $\sigmav$ assuming no pre-existing anisotropy and 
	$\sigmav_\mathrm{syst}$ assuming the $N_{90}$ changed by 
	the systematic effects.}
	\label{tab:bgd_syst}
	\centering
	\begin{tabular}{ l | r }
		\hline\hline
		uncertainty & $\frac{\langle\sigma_Av\rangle_\mathrm{syst}-\langle\sigma_Av\rangle_\mathrm{base}}{\langle\sigma_Av\rangle_\mathrm{base}} [\%]$ \\
		\hline
		zenith acceptance &  $<4.3$ \\
		sky exposure & $\pm2.2$ \\
		cosmic ray anisotropy & $\pm5.4$ \\
		\hline\hline
	\end{tabular}
\end{table}

As the background expectation is generated from scrambled 
experimental events in RA, systematic effects can only be caused by 
pre-existing anisotropies. 
Such an anisotropy can arise from the zenith-dependent acceptance of the detector,
the zenith-dependent variation of the atmospheric neutrino flux or the detector exposure.
There is also the possibility of an anisotropy in the atmospheric neutrino flux, caused by the
cosmic-ray anisotropy which has been measured by Milagro~\cite{MilagroCRAnisotropy}, 
TUNKA~\cite{sveshnikova_spectrum_2013}, ARGO-YBJ~\cite{bartoli_medium_2013} and IceCube~\cite{abbasi_measurement_2010}.

The systematic uncertainty on the self-annihilation cross-section caused by zenith-dependent 
uncertainties is very small as a result of the fact that the coefficients corresponding to pure zenith(declination)-dependent spherical harmonics are not included in the test statistic. 
In order to study the influence of the zenith structure that arises from the acceptance of the detector and the  variation of the atmospheric neutrino flux, pseudo-experiments 
were generated using events according to a histogram of experimental zenith values and not using the experimental data directly. 
To generate steeper and flatter zenith-spectra, the bin-contents of the histogram 
are changed by raising the outer most left bin by $25\,\%$ and decreasing the 
outer most right bin by $25\,\%$. The bins in between are raised or decreased according to a linear interpolation between $+25\,\%$ and $-25\,\%$.
The uncertainties on the 
zenith-spectrum are on the order of 5\,\%. However to study this 
effect and not be limited by statistics the slope of the zenith-spectrum was 
changed by $\pm50\,\%$ resulting in a large overestimation of the effect. 
Based on these pseudo-experiments the median sensitivity on $\sigmav$ 
was calculated. This results in a conservative upper limit on the effect of zenith-dependent uncertainties (see Table~\ref{tab:bgd_syst}). 
 
The up-time of the IceCube-detector is of the order of 98\,\%, however due 
to high quality criteria in the data selection the used data correspond to 91\,\% up-time.  
The geometry of the detector is almost symmetric in azimuth, and thus 
the exposure of each direction in the sky is nearly constant. 
However due to short down-times and a non flat azimuth acceptance an 
anisotropy of $0.02\,\%$ in the data sample ($\sim10$ events) can 
occur. In the worst case this anisotropy can mimic a (anti-)signal 
and thus result in a small systematic effect on the median sensitivity $\sigmav$ 
(see Table~\ref{tab:bgd_syst}).

Milagro, ARGO-YBJ and TUNKA have observed an anisotropy in cosmic rays at few hundred $\GeV$ to $\EeV$ 
energies of primary particles in the Northern Hemisphere~\cite{MilagroCRAnisotropy, sveshnikova_spectrum_2013, bartoli_medium_2013}. 
Because the experimental sky map is dominated by atmospheric 
neutrinos, that were produced in air-showers initiated by cosmic rays an 
analogous anisotropy is expected in atmospheric neutrinos. Therefore, 
pseudo-experiments were generated that allow for an anisotropy as 
parameterized in~\cite{MilagroCRAnisotropy}. The uncertainty on 
the median sensitivity on $\sigmav$ is given in 
Table~\ref{tab:bgd_syst}.

\begin{table}
	\caption{Relative systematic uncertainties on $\sigmav$ 
	resulting from uncertainties in the detection efficiency. 
	Because the detection efficiency is energy-dependent, the 
	uncertainties are given in dependence of annihilation channel and
	the mass of the DM particle $m_\chi$. $\sigmav_\mathrm{base}$ 
	denotes $\sigmav$ assuming the baseline effective area and 
	$\sigmav_\mathrm{syst}$ assuming the effective area changed by 
	the systematic effects.} 
	\label{tab:sig_syst} 
	\centering 
	\begin{tabular}{c|c|c|c|c} 
		\hline\hline
		$m_\chi \,[\GeV]$ & \multicolumn{4}{c}{ $\frac{\langle\sigma_Av\rangle_\mathrm{syst}-\langle\sigma_Av\rangle_\mathrm{base}}{\langle\sigma_Av\rangle_\mathrm{base}} [\%]$} \\
		 & $b\bar{b}$ & $W^+W^-$ & $\mu^+\mu^-$ & $\nu\bar{\nu}$ \\ 
		\hline 
$100$ & $\pm89$ & $\pm30$ & $\pm32$ & $\pm33$ \\ 
$200$ & $\pm34$ & $\pm28$ & $\pm30$ & $\pm29$ \\ 
$300$ & $\pm26$ & $\pm25$ & $\pm27$ & $\pm24$ \\ 
$400$ & $\pm27$ & $\pm22$ & $\pm24$ & $\pm15$ \\ 
$500$ & $\pm27$ & $\pm20$ & $\pm22$ & $\pm18$ \\ 
$600$ & $\pm26$ & $\pm19$ & $\pm20$ & $\pm22$ \\ 
$700$ & $\pm25$ & $\pm18$ & $\pm19$ & $\pm15$ \\ 
$800$ & $\pm25$ & $\pm18$ & $\pm19$ & $\pm17$ \\ 
$900$ & $\pm24$ & $\pm18$ & $\pm18$ & $\pm18$ \\ 
$1000$ & $\pm23$ & $\pm18$ & $\pm18$ & $\pm14$ \\ 
$2000$ & $\pm20$ & $\pm15$ & $\pm16$ & $\pm12$ \\ 
$3000$ & $\pm19$ & $\pm14$ & $\pm14$ & $\pm16$ \\ 
$4000$ & $\pm18$ & $\pm13$ & $\pm13$ & $\pm15$ \\ 
$5000$ & $\pm18$ & $\pm12$ & $\pm12$ & $\pm13$ \\ 
$10000$ & $\pm16$ & $\pm10$ & $\pm11$ & $\pm9$ \\ 
$20000$ & $\pm14$ & $\pm9$ & $\pm7$ & $\pm12$ \\ 
$30000$ & $\pm13$ & $\pm10$ & $\pm8$ & $\pm10$ \\ 
$50000$ & $\pm12$ & $\pm7$ & $\pm7$ & $\pm24$ \\ 
$70000$ & $\pm11$ & $\pm6$ & $\pm4$ & $\pm13$ \\ 
$100000$ & $\pm10$ & $\pm6$ & $\pm3$ & $\pm26$ \\ 
		\hline\hline
	\end{tabular} 
\end{table} 

Systematic uncertainties on the signal efficiency can be expressed by 
uncertainties in the effective area. Because the effective area 
depends on energy, the resulting systematic uncertainties on $\sigmav$
depend on the assumed energy-spectrum and, thus, on the 
annihilation channel and the mass of the dark matter particle.

The main uncertainties of the detection efficiency arise from the 
optical efficiency of the DOMs and from the optical properties of the antarctic ice, 
described by the absorption and scattering length.
The influence of these effects on the effective area was determined by
a full detector simulation where the nominal values of the DOM efficiency
and the absorption and scattering lengths were changed by $\pm10\,\%$~\cite{PMTCalibration,IceCalibration}.
The uncertainties on the effective area were further propagated to 
uncertainties on $\sigmav$, which depends on the dark matter particle mass 
$m_\chi$ and the annihilation channel. The resulting uncertainties 
are listed in Table~\ref{tab:sig_syst}. 
They typically lie in the range 15\,\%-30\,\%, and they are the dominating uncertainties of this analysis.

The sensitivities as obtained from the different halo profiles using 
best-fit parameters differ by about $6\,\%$. This is smaller than the 
uncertainty that arises from uncertainties on the profile fit values.
The dominant contribution comes from the local dark matter density, 
and corresponds to an uncertainty on the sensitivity of up to $50\,\%$.
In the 
following the dependency of the assumed model is not treated as a 
systematic uncertainty, but as model uncertainty, and thus the 
experimental result will be interpreted for each of the different halo profiles, and 
benchmark annihilation channels, respectively.
 

\section{Experimental Results}\label{sec:results}

\begin{figure}
	\includegraphics[width=1.\linewidth]{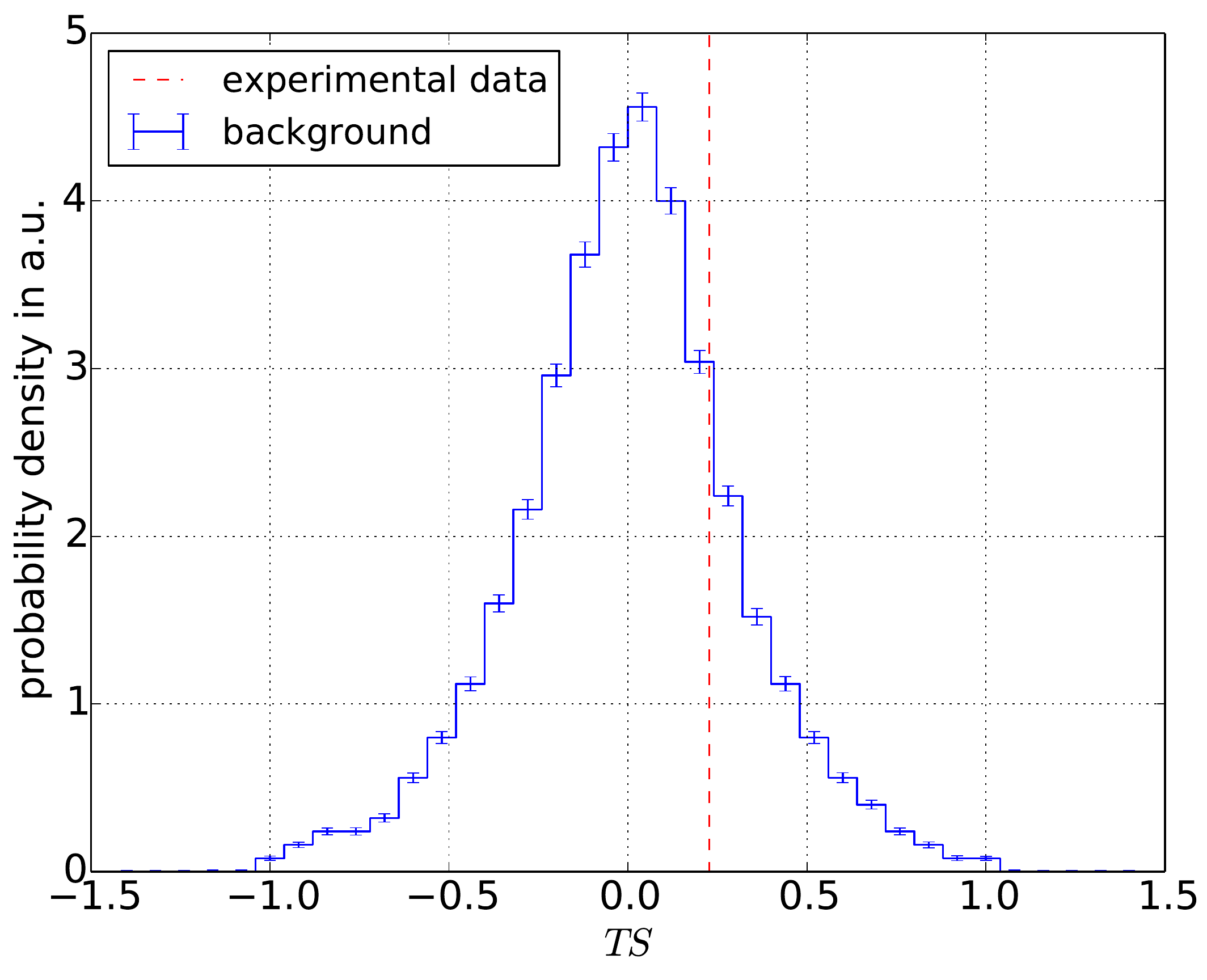} 
	\caption{$TS$ distribution for background expectation (solid), 
	and the observed value $TS_{\mathrm{exp}} = 0.23$ 
	(dashed). The error bars on the background distribution reflect the 
	statistical precision arising from the finite number of pseudo-experiments realized.}
	\label{fig:D2Result}
\end{figure}
\begin{figure}
	\includegraphics[width=1.\linewidth]{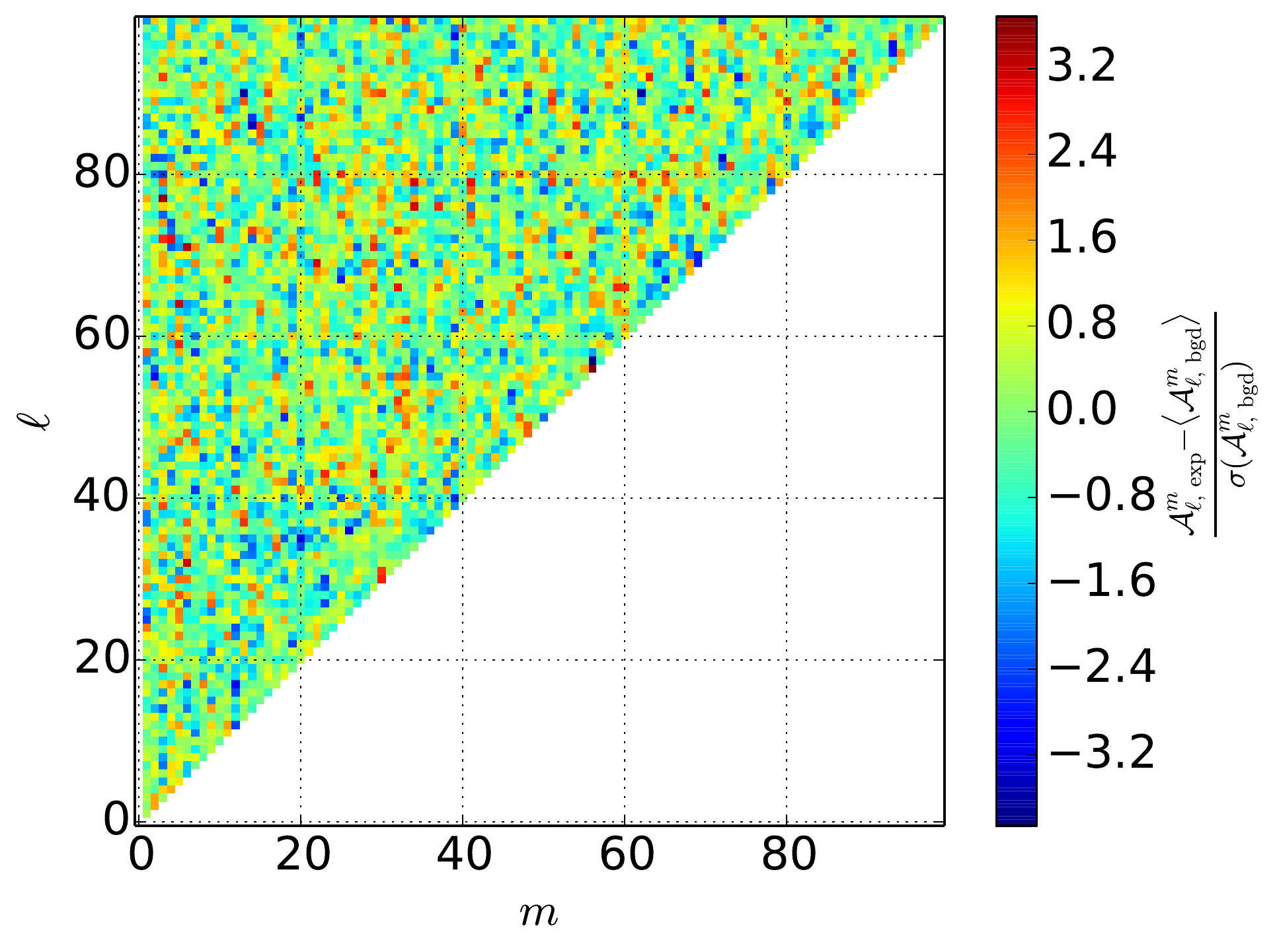} 
	\caption{Deviation of experimental 
	projected expansion coefficients from background expectation, 
	normalized to standard deviation of background coefficients in 
	the $\ell$-$m$-plane. No significant excess can be seen.} 
	\label{fig:exp_Alm_plain}
\end{figure}

This analysis was performed blind, meaning it was developed by using 
pseudo-experiments only. After the analysis procedure was optimized 
and fixed, the data were unblinded. The experimental 
sky map has a test statistic value of $TS_{\mathrm{exp}} = 0.23$. 
The probability of a larger experimental value in the 
background-only case is $22\,\%$ and thus the result is compatible 
with the background-only hypothesis. The observation is an over-fluctuation corresponding to 
$0.8\sigma$, where $\sigma$ is the standard deviation of the 
background expectation of the test statistic. Note that the 
test statistic can not be approximated by a Gaussian 
due to larger tails. The experimental value and the background 
expectation of the test statistic are shown in Figure~\ref{fig:D2Result}.

Figure~\ref{fig:exp_Alm_plain} shows the deviation of the 
experimental expansion coefficients from the background expectation, 
normalized to the standard deviation of the background coefficients. 
These values correspond to the last term in Equation~\ref{eq:test_statistic} 
without the square. Also here no significant 
deviation can be seen.

As no signal was observed, upper limits on the number of signal events 
in the sample, $N_\mathrm{UL}$, were calculated at a 90\,\% C.L. 
following the approach of Feldman and Cousins~\cite{FeldmanCousins}. 
In order to calculate the confidence belt, 25000 pseudo-experiments for different $N_\mathrm{sig}$ 
were generated respectively. Due to limited computational resources 
the pseudo-experiments were not generated for each $N_\mathrm{sig}$, 
but for signal contributions $N_\mathrm{sim,i}$ with a step-size of 
$\Delta_\mathrm{sim}=25$ events. The test statistic distribution was 
interpolated for the remaining $N_\mathrm{sig}$, using a Gaussian, 
$p_\mathrm{gaus}$, with mean $\mu=N_\mathrm{sig}$ and standard 
deviation $\sigma=\sqrt{N_\mathrm{sig}}$ . The interpolated test statistic 
distributions are given by
\begin{equation}
	TS(N_\mathrm{sig}) = 
		\sum_i
			TS(N_\mathrm{sim,i}) \cdot
			\int_{N_\mathrm{sim,i}-\Delta_\mathrm{sim}/2}^{N_\mathrm{sim,i}+\Delta_\mathrm{sim}/2}
			p_\mathrm{gaus}(N) \mathrm{d}N
	\, ,
	\label{eq:D2_interpolation}
\end{equation}
where $i$ runs over all generated test statistics. 
The result of the pseudo-experiments (number of signal neutrinos) was 
smeared by a Gaussian with width corresponding to the systematic uncertainties, 
as described in Section~\ref{sec:systematics}. Systematic errors, including the uncertainty on the
effective area, are thus included in the effective upper limits on the
number of events, listed in Table~\ref{tab:N_limit}. These can be directly translated to
limits on $\sigmav$ using equation~\ref{eq:sigmav_90}.

\begin{table}
	\caption{Effective 90\,\% C.L. Feldman-Cousins limit on the number of 
	signal events in the data set, $N_\mathrm{UL}$, for different 
	halo profiles. The values include systematics and can be directly translated to limits
on $\sigmav$ using equation~\ref{eq:sigmav_90}.}
	\label{tab:N_limit}
	\centering
	\begin{tabular}{ l | r }
		\hline\hline
		Halo Profile & $N_\mathrm{UL}$ \\
		\hline
NFW & $949$ \\Burkert & $1014$\\		\hline\hline
	\end{tabular}
\end{table}

\begin{figure}
	\includegraphics[width=1.\linewidth]{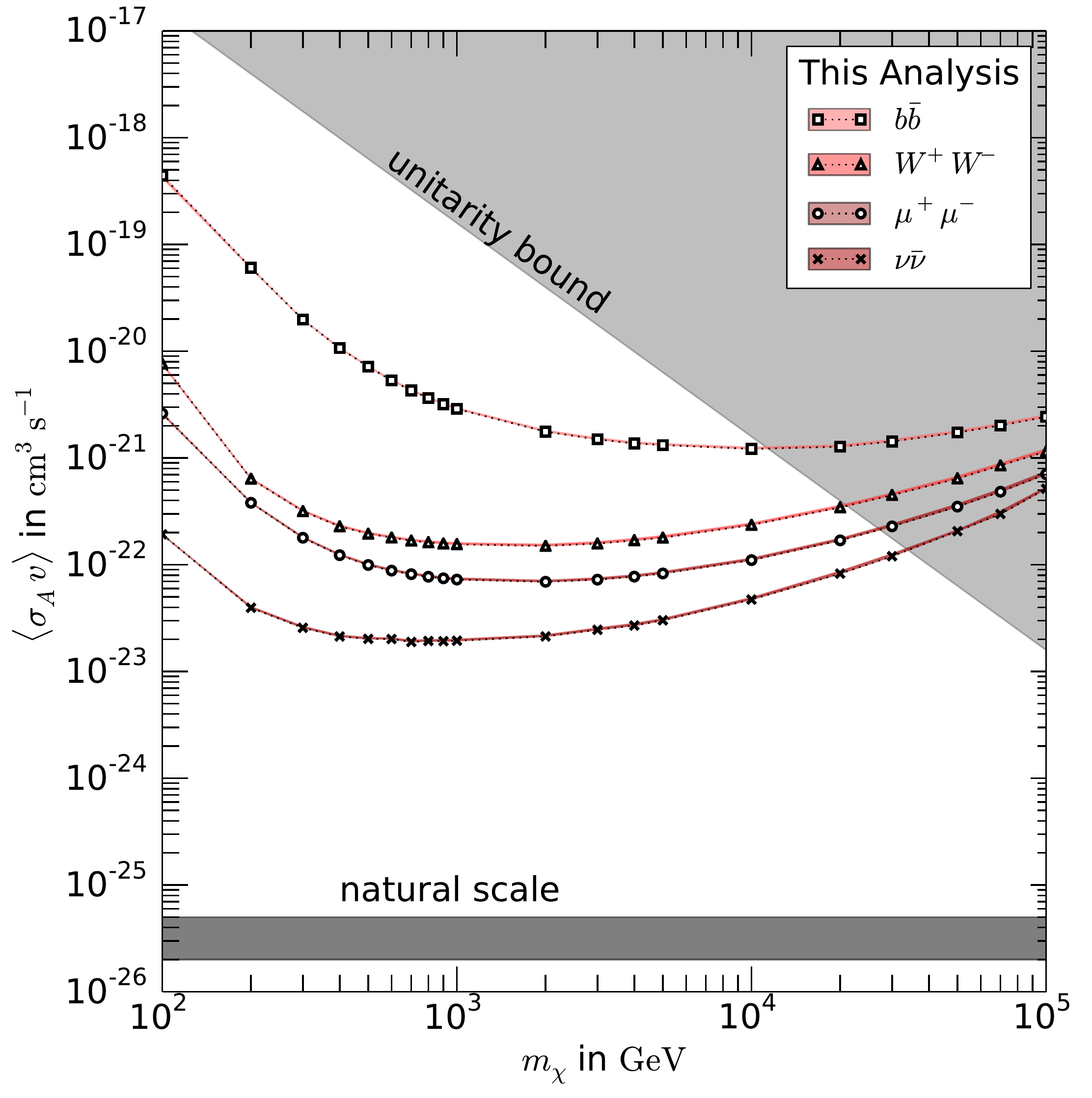} 
	\caption{
	Exclusion limits on dark matter self-annihilation cross-section from 
	this analysis at 90\,\% 
	C.L.. The baseline limit curves are 
	calculated for the NFW profile. The model-dependence has 
	been estimated from the Burkert profile and are shown as bands, 
	which are very narrow and thus hard to see. The gray band describes the natural scale 
	if all dark matter consists of WIMPs and the gray upper region is excluded by the unitarity bound~\cite{WIMPMass2}.}
	\label{fig:LimitPlot}
\end{figure}

By using equation (\ref{eq:sigmav_90}) the limit on the signal events $N_\mathrm{UL}$ can be interpreted in terms of a 
limit on the self-annihilation cross-section $\sigmav_\mathrm{UL}$.
The resulting limits are shown in Figure~\ref{fig:LimitPlot} as function of 
$m_\chi$ and for the different benchmark annihilation channels.  
The limits are also listed in Table~\ref{tab:lim_table}. 
In correspondence to the experimental exclusion limit it is possible to calculate the average upper limit, 
which gives the mean expected exclusion limit in case of no signal~\cite{averaged_upper_limit}. 
The average effective upper limit on the number of signal neutrinos in case of an $W^+W^-$ 
annihilation channel and a dark matter particle mass of $600\,\GeV$ is $\langle N_{\mathrm{UL}}\rangle = 747$. 
Note that the average upper limit is more stringent by 10\%-24\% than the resulting exclusion limits depending on the halo profile annihilation channel and dark matter particle mass.

\begin{table}
	\caption{Limit on the self-annihilation cross-section $\sigmav$ 
	for different annihilation channels, halo profiles and 
	DM-particle masses $m_\chi$.}
	\label{tab:lim_table} 
	\centering
	\begin{tabular}{c|c|c|c|c}
\hline \hline$m_\chi \,[\GeV]$ & \multicolumn{4}{c}{ $\langle \sigma_A v\rangle \left[ \mathrm{cm}^{3}\mathrm{s}^{-1} \right]$ assuming Burkert profile} \\& $b\bar{b}$ & $W^+W^-$ & $\mu^+\mu^-$ & $\nu\bar{\nu}$ \\\hline
$100$ & $4.2\cdot10^{-19}$ & $7.6\cdot10^{-21}$ & $2.6\cdot10^{-21}$ & $1.9\cdot10^{-22}$ \\ 
$200$ & $6.0\cdot10^{-20}$ & $6.5\cdot10^{-22}$ & $3.8\cdot10^{-22}$ & $4.0\cdot10^{-23}$ \\ 
$300$ & $2.0\cdot10^{-20}$ & $3.2\cdot10^{-22}$ & $1.8\cdot10^{-22}$ & $2.6\cdot10^{-23}$ \\ 
$400$ & $1.1\cdot10^{-20}$ & $2.3\cdot10^{-22}$ & $1.3\cdot10^{-22}$ & $2.2\cdot10^{-23}$ \\ 
$500$ & $7.3\cdot10^{-21}$ & $2.0\cdot10^{-22}$ & $1.0\cdot10^{-22}$ & $2.1\cdot10^{-23}$ \\ 
$600$ & $5.4\cdot10^{-21}$ & $1.8\cdot10^{-22}$ & $9.0\cdot10^{-23}$ & $2.1\cdot10^{-23}$ \\ 
$700$ & $4.4\cdot10^{-21}$ & $1.7\cdot10^{-22}$ & $8.4\cdot10^{-23}$ & $2.0\cdot10^{-23}$ \\ 
$800$ & $3.7\cdot10^{-21}$ & $1.7\cdot10^{-22}$ & $7.9\cdot10^{-23}$ & $2.0\cdot10^{-23}$ \\ 
$900$ & $3.3\cdot10^{-21}$ & $1.6\cdot10^{-22}$ & $7.7\cdot10^{-23}$ & $2.0\cdot10^{-23}$ \\ \hdashline
$1000$ & $2.9\cdot10^{-21}$ & $1.6\cdot10^{-22}$ & $7.5\cdot10^{-23}$ & $2.0\cdot10^{-23}$ \\ 
$2000$ & $1.8\cdot10^{-21}$ & $1.5\cdot10^{-22}$ & $7.1\cdot10^{-23}$ & $2.2\cdot10^{-23}$ \\ 
$3000$ & $1.5\cdot10^{-21}$ & $1.6\cdot10^{-22}$ & $7.5\cdot10^{-23}$ & $2.5\cdot10^{-23}$ \\ 
$4000$ & $1.4\cdot10^{-21}$ & $1.8\cdot10^{-22}$ & $8.0\cdot10^{-23}$ & $2.8\cdot10^{-23}$ \\ 
$5000$ & $1.4\cdot10^{-21}$ & $1.9\cdot10^{-22}$ & $8.6\cdot10^{-23}$ & $3.1\cdot10^{-23}$ \\ \hdashline
$10000$ & $1.3\cdot10^{-21}$ & $2.5\cdot10^{-22}$ & $1.1\cdot10^{-22}$ & $4.9\cdot10^{-23}$ \\ 
$20000$ & $1.3\cdot10^{-21}$ & $3.6\cdot10^{-22}$ & $1.8\cdot10^{-22}$ & $8.7\cdot10^{-23}$ \\ 
$30000$ & $1.5\cdot10^{-21}$ & $4.7\cdot10^{-22}$ & $2.4\cdot10^{-22}$ & $1.3\cdot10^{-22}$ \\ 
$50000$ & $1.8\cdot10^{-21}$ & $6.7\cdot10^{-22}$ & $3.7\cdot10^{-22}$ & $2.1\cdot10^{-22}$ \\ 
$70000$ & $2.1\cdot10^{-21}$ & $8.9\cdot10^{-22}$ & $5.1\cdot10^{-22}$ & $3.2\cdot10^{-22}$ \\ 
$100000$ & $2.5\cdot10^{-21}$ & $1.2\cdot10^{-21}$ & $7.4\cdot10^{-22}$ & $5.3\cdot10^{-22}$ \\ 
\hline \hline$m_\chi \,[\GeV]$ & \multicolumn{4}{c}{ $\langle \sigma_A v\rangle \left[ \mathrm{cm}^{3}\mathrm{s}^{-1} \right]$ assuming NFW profile} \\& $b\bar{b}$ & $W^+W^-$ & $\mu^+\mu^-$ & $\nu\bar{\nu}$ \\\hline 
$100$ & $4.4\cdot10^{-19}$ & $7.6\cdot10^{-21}$ & $2.6\cdot10^{-21}$ & $1.9\cdot10^{-22}$ \\ 
$200$ & $6.1\cdot10^{-20}$ & $6.4\cdot10^{-22}$ & $3.8\cdot10^{-22}$ & $4.0\cdot10^{-23}$ \\ 
$300$ & $2.0\cdot10^{-20}$ & $3.2\cdot10^{-22}$ & $1.8\cdot10^{-22}$ & $2.6\cdot10^{-23}$ \\ 
$400$ & $1.1\cdot10^{-20}$ & $2.3\cdot10^{-22}$ & $1.2\cdot10^{-22}$ & $2.1\cdot10^{-23}$ \\ 
$500$ & $7.2\cdot10^{-21}$ & $2.0\cdot10^{-22}$ & $1.0\cdot10^{-22}$ & $2.0\cdot10^{-23}$ \\ 
$600$ & $5.3\cdot10^{-21}$ & $1.8\cdot10^{-22}$ & $8.8\cdot10^{-23}$ & $2.0\cdot10^{-23}$ \\ 
$700$ & $4.3\cdot10^{-21}$ & $1.7\cdot10^{-22}$ & $8.2\cdot10^{-23}$ & $1.9\cdot10^{-23}$ \\ 
$800$ & $3.6\cdot10^{-21}$ & $1.6\cdot10^{-22}$ & $7.7\cdot10^{-23}$ & $1.9\cdot10^{-23}$ \\ 
$900$ & $3.2\cdot10^{-21}$ & $1.6\cdot10^{-22}$ & $7.5\cdot10^{-23}$ & $1.9\cdot10^{-23}$ \\ \hdashline
$1000$ & $2.9\cdot10^{-21}$ & $1.6\cdot10^{-22}$ & $7.3\cdot10^{-23}$ & $1.9\cdot10^{-23}$ \\ 
$2000$ & $1.8\cdot10^{-21}$ & $1.5\cdot10^{-22}$ & $6.9\cdot10^{-23}$ & $2.1\cdot10^{-23}$ \\ 
$3000$ & $1.5\cdot10^{-21}$ & $1.6\cdot10^{-22}$ & $7.3\cdot10^{-23}$ & $2.5\cdot10^{-23}$ \\ 
$4000$ & $1.4\cdot10^{-21}$ & $1.7\cdot10^{-22}$ & $7.7\cdot10^{-23}$ & $2.7\cdot10^{-23}$ \\ 
$5000$ & $1.3\cdot10^{-21}$ & $1.8\cdot10^{-22}$ & $8.3\cdot10^{-23}$ & $3.0\cdot10^{-23}$ \\ \hdashline
$10000$ & $1.2\cdot10^{-21}$ & $2.4\cdot10^{-22}$ & $1.1\cdot10^{-22}$ & $4.7\cdot10^{-23}$ \\ 
$20000$ & $1.3\cdot10^{-21}$ & $3.4\cdot10^{-22}$ & $1.7\cdot10^{-22}$ & $8.3\cdot10^{-23}$ \\ 
$30000$ & $1.4\cdot10^{-21}$ & $4.5\cdot10^{-22}$ & $2.3\cdot10^{-22}$ & $1.2\cdot10^{-22}$ \\ 
$50000$ & $1.7\cdot10^{-21}$ & $6.4\cdot10^{-22}$ & $3.5\cdot10^{-22}$ & $2.1\cdot10^{-22}$ \\ 
$70000$ & $2.0\cdot10^{-21}$ & $8.5\cdot10^{-22}$ & $4.8\cdot10^{-22}$ & $3.0\cdot10^{-22}$ \\ 
$100000$ & $2.4\cdot10^{-21}$ & $1.2\cdot10^{-21}$ & $7.1\cdot10^{-22}$ & $5.2\cdot10^{-22}$ \\ 
		\hline
		\hline
	\end{tabular}
\end{table}


\section{Discussion} \label{sec:discussion}

\begin{figure}
	\includegraphics[width=1.\linewidth]{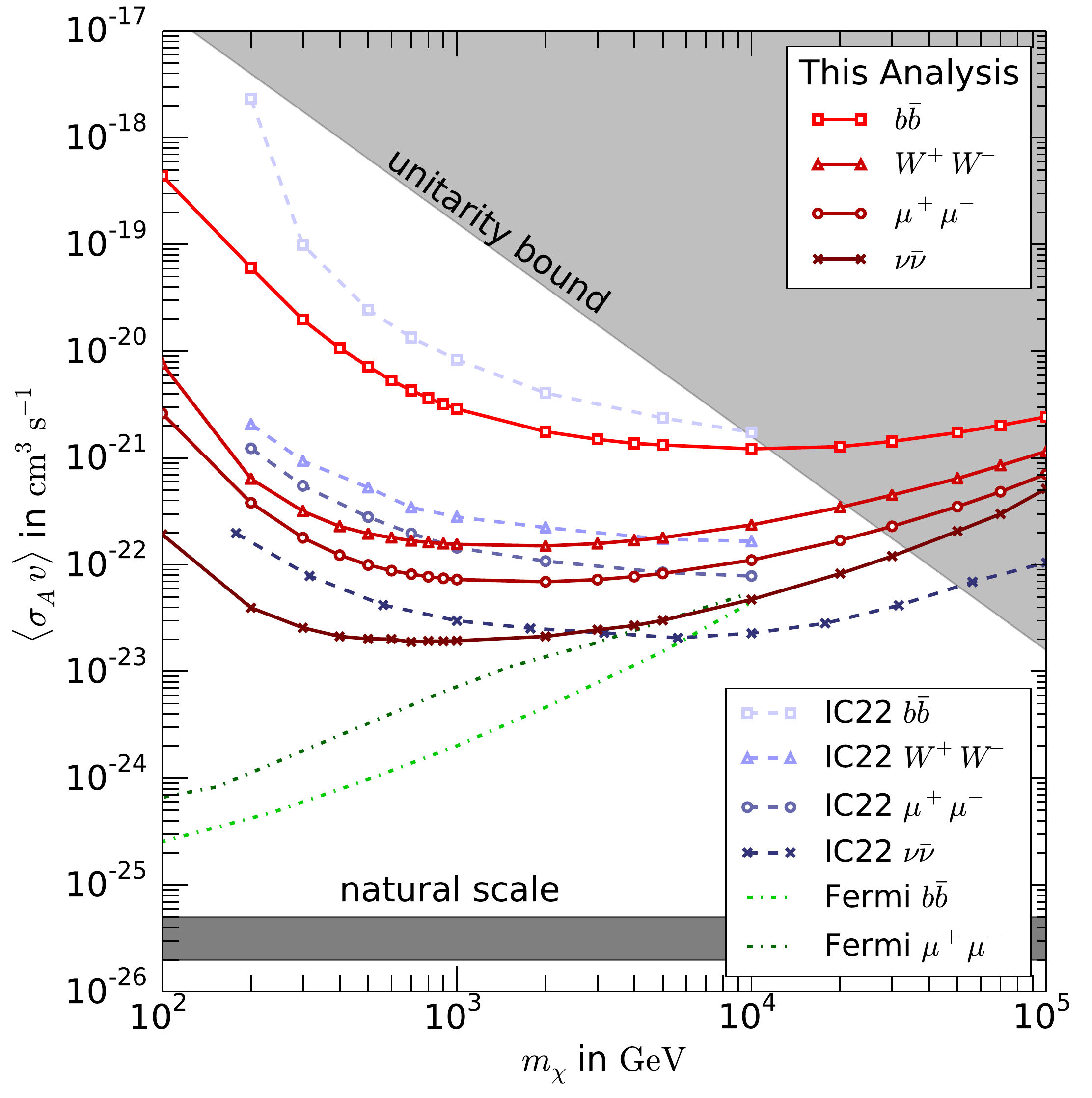} 
	\caption{ 
	Exclusion limits on dark matter self-annihilation cross-section from outer Galactic halo searches only.
	The results of this analysis and 
	exclusion limits from the predecessor analysis of IceCube-22~\cite{IC22} (both 90\,\% 
	C.L.) and from the Fermi-LAT~\cite{fermi} ($3\sigma$ C.L.) are shown. 
	The baseline limit curves are 
	calculated for the NFW profile, however different normalization 
	parameter are used. For reasons of comparison the limits from 
	IC22 and Fermi are rescaled to the local density of 
	$\rho_{\mathrm{SC}}=0.47\,\GeV/\cm^3$ that is used in this 
	analysis. The gray band describes the natural scale if all dark 
	matter consists of WIMPs and the gray upper region is excluded 
	by the unitarity bound~\cite{WIMPMass2}.}
	\label{fig:Comp_all}
\end{figure}

Compared to the predecessor analysis of IC22 data using a 
cut-and-count based method~\cite{IC22}, the effective area 
increases by more than an order of magnitude in the low energy 
region ($\sim100\,\GeV$) but just a factor of about 3 at high 
energies ($\sim10\,\TeV$) in the relevant zenith region. The larger 
gain in effective area for low dark matter masses is caused by DeepCore, the 
low-energy extension of IceCube, which was not implemented in IC22, 
but was already in operation in IC79. The lager gain in the 
effective area at low energies causes an increase in sensitivity of 
more than an order of magnitude at these energies. However due to a 
much larger sample size, caused 
by the large increase in the number of low energy events, 
and just a slight increase in effective area at high energies, there is 
just a small gain in sensitivity for high dark matter masses.
As this analysis measures an over-fluctuation and the IC22 
analysis has measured a under-fluctuation the exclusion 
limits of the IC22 analysis are more stringent for high dark matter masses of 
a few $\TeV$. However the limits in the low-mass region of 
$100\,\GeV$ are still an order of magnitude more stringent due to the larger increase in sensitivity.
For comparison the exclusion limits (90\,\% C.L.) of the predecessor 
analysis are shown in Figure~\ref{fig:Comp_all}. Note that in the 
predecessor analysis a NFW profile was assumed, but with a local density of 
$0.3\,\GeV/\mathrm{cm}^3$, while here $0.47\,\GeV/\mathrm{cm}^3$ is 
assumed. For reasons of comparison the limits in Figure~\ref 
{fig:Comp_all} have been rescaled to the local density used in this 
analysis. 
 
Furthermore, exclusion limits of an outer Galactic halo analysis by 
Fermi-LAT~\cite{fermi} are also shown in Figure~\ref{fig:Comp_all} 
for annihilation into $b\bar{b}$ and $\mu^+\mu^-$. This analysis has 
measured the $\gamma$-ray emission along the Galactic plane in a 
window of $\pm15^\circ$, whereas the central $\pm5^\circ$ are 
excluded. In~\cite{fermi} a NFW profile was assumed, but with a 
local density of $0.43\,\GeV/\mathrm{cm}^3$, while here 
$0.47\,\GeV/\mathrm{cm}^3$ is assumed. For reasons of comparison the 
limits in Figure~\ref{fig:Comp_all} have been rescaled to the local 
density used in this analysis. It can be seen that for hard neutrino 
channels ($\mu^+\mu^-$) the exclusion limits of this analysis  
come close to the outer Galactic halo limits of Fermi. 

\begin{figure}
	\includegraphics[width=1.\linewidth]{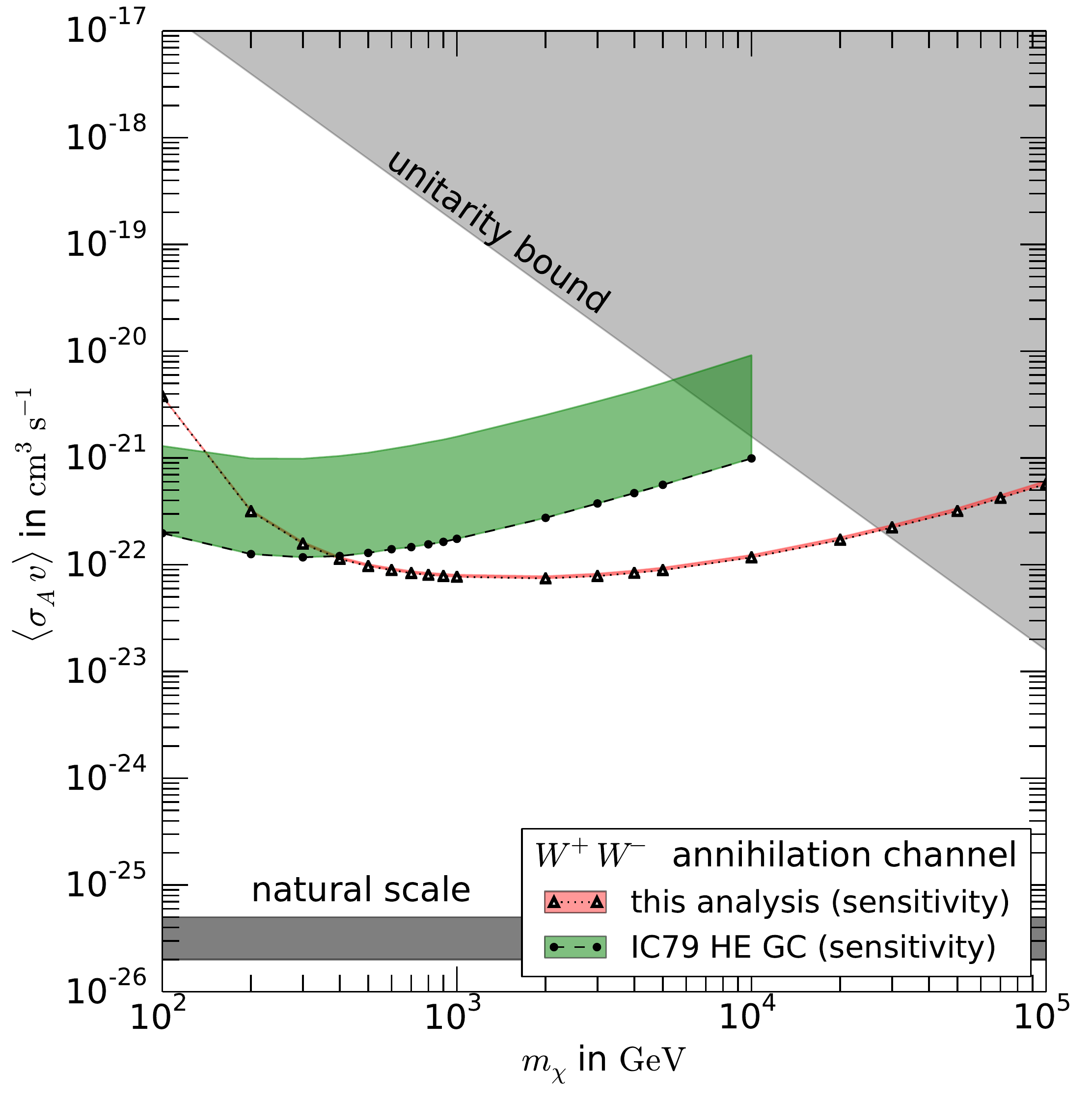} 
	\caption{Median sensitivity on dark matter self-annihilation 
	cross-section assuming annihilation in $W^+W^-$ for this 
	analysis  
	and for the IceCube-79 high energy Galactic Center analysis 
	(IC79 HE GC)~\cite{Aartsen:2013mla}. The baseline limit curves are calculated for the NFW 
	profile (markers). The model-dependence (bands) has been estimated from 
	the Burkert profile. The gray band describes the 
	natural scale if all dark matter consists of WIMPs and the gray upper region is excluded by the unitarity bound~\cite{WIMPMass2}.}
	\label{fig:Comp_GC}
\end{figure}

The most stringent exclusion limits from $\gamma$-ray telescopes in the 
energy-range of this analysis are set by H.E.S.S~\cite 
{PhysRevLett.106.161301}, with an analysis focusing on the Galactic 
Center, and Fermi~\cite{collaboration_constraining_2011} with an 
analysis focusing at dwarf galaxies. These limits are about two 
orders of magnitude more stringent. However it is important to note, 
that the systematic uncertainties for $\gamma$-ray and neutrino 
telescopes are of very different nature. Also $\gamma$ 
telescopes have the highest sensitivity for the soft channel and 
vice versa, smallest sensitivity for the hard channels.

The halo profile dependencies in this analysis are very small 
compared to the Galactic Center analysis of IceCube. This can be seen by comparing this 
analysis with a Galactic Center search, that focuses on the central part of the galaxy. 
The sensitivity for the $W^+W^-$ annihilation channel of the 
IceCube-79 Galactic Center analysis described in~\cite{Aartsen:2013mla} and the sensitivity of this analysis are compared 
in Figure~\ref{fig:Comp_GC}. The bands represent the model 
uncertainties determined from Burkert and NFW profile, whereas NFW is 
used as baseline. It is clearly visible that the halo profile 
uncertainties are much smaller for a halo analysis (compare Figure~\ref{fig:JofPsi}), while the overall sensitivity of the two approaches are remarkably similar.


\section{Conclusions}\label{sec:conclusion}

We have presented a competitive analysis technique to search for characteristic 
anisotropies by using a multipole expansion of the neutrino 
arrival direction sky map. It was found that the 
multipole analysis is a sensitive and robust analysis method, that has 
the feature to reduce systematic uncertainties in an easy way.

We applied the analysis to one year of data taken with the IceCube detector in its the nearly completed detector configuration. 
The search for a neutrino flux, resulting from dark matter annihilation, has 
found no significant deviation from the background expectation. 
Exclusion limits on the self-annihilation cross-section $\sigmav$ 
were calculated approaching $1.9\cdot 
10^{-23}\,\textrm{cm}^3\textrm{s}^{-1}$. The resulting exclusion 
limits are more stringent than the predecessor analysis of IC22 data 
in a wide parameter range. Furthermore the extracted limits  
come close to limits from $\gamma$-experiments, that also focus on 
the outer Galactic halo, for hard annihilation channels and large dark matter 
masses. 
The presented analysis, focusing on the Galactic halo, is very robust 
against halo profile uncertainties compared to analyses targeting the 
Galactic Center or dwarf spheroidal galaxies (compare Figure~\ref{fig:JofPsi}).


\begin{acknowledgements} 

We acknowledge the support from the following agencies:
U.S. National Science Foundation-Office of Polar Programs,
U.S. National Science Foundation-Physics Division,
University of Wisconsin Alumni Research Foundation,
the Grid Laboratory Of Wisconsin (GLOW) grid infrastructure at the University of Wisconsin - Madison, the Open Science Grid (OSG) grid infrastructure;
U.S. Department of Energy, and National Energy Research Scientific Computing Center,
the Louisiana Optical Network Initiative (LONI) grid computing resources;
Natural Sciences and Engineering Research Council of Canada,
WestGrid and Compute/Calcul Canada;
Swedish Research Council,
Swedish Polar Research Secretariat,
Swedish National Infrastructure for Computing (SNIC),
and Knut and Alice Wallenberg Foundation, Sweden;
German Ministry for Education and Research (BMBF),
Deutsche Forschungsgemeinschaft (DFG),
Helmholtz Alliance for Astroparticle Physics (HAP),
Research Department of Plasmas with Complex Interactions (Bochum), Germany;
Fund for Scientific Research (FNRS-FWO),
FWO Odysseus programme,
Flanders Institute to encourage scientific and technological research in industry (IWT),
Belgian Federal Science Policy Office (Belspo);
University of Oxford, United Kingdom;
Marsden Fund, New Zealand;
Australian Research Council;
Japan Society for Promotion of Science (JSPS);
the Swiss National Science Foundation (SNSF), Switzerland;
National Research Foundation of Korea (NRF);
Danish National Research Foundation, Denmark (DNRF)

\end{acknowledgements}


\hyphenation{Post-Script Sprin-ger}

\end{document}